\documentclass[twocolumn,twocolappendix]{aastex63}
\usepackage{booktabs}
\usepackage{threeparttable}
\usepackage{tabularx}
\usepackage{multirow}
\usepackage{graphicx}
\usepackage[caption=false]{subfig}

\usepackage{amsmath}

\newcommand\agel{\texttt{AGEL}}
\newcommand\reb{\agel0014}
\newcommand\rs{\agel1323}
\newcommand\alf{\texttt{alf}}

\shortauthors{Z. Zhuang et al.}

\graphicspath{{./}{figures/}}

\begin{document}

\title{A Glimpse of the Stellar Populations and Elemental Abundances of Gravitationally Lensed, Quiescent Galaxies at $z\gtrsim 1$ with Keck Deep Spectroscopy
}
\shorttitle{Stellar Abundances of High-$z$ Lensed, Quiescent Galaxies }

\correspondingauthor{Zhuyun Zhuang}
\email{zzhuang@astro.caltech.edu}

\author[0000-0002-1945-2299]{Zhuyun Zhuang}
\affiliation{Department of Astronomy, California Institute of Technology, 1200 E. California Blvd., MC 249-17, Pasadena, CA 91125, USA}

\author[0000-0003-4570-3159]{Nicha Leethochawalit}
\affiliation{National Astronomical Research Institute of Thailand (NARIT), MaeRim, Chiang Mai, 50180, Thailand}
\affiliation{School of Physics, The University of Melbourne, Parkville, VIC 3010, Australia}
\affiliation{ARC Centre of Excellence for All Sky Astrophysics in 3 Dimensions (ASTRO 3D), Australia}

\author[0000-0001-6196-5162]{Evan N.\ Kirby}
\affiliation{Department of Physics and Astronomy, University of Notre Dame, 225 Nieuwland Science Hall, Notre Dame, IN 46556, USA}

\author[0000-0002-8987-7401]{J.\ W.\ Nightingale}
\affiliation{Centre for Extragalactic Astronomy, Department of Physics, Durham University, South Road, Durham, DH1 3LE, UK}

\author[0000-0002-4834-7260]{Charles C. Steidel}
\affiliation{Department of Astronomy, California Institute of Technology, 1200 E. California Blvd., MC 249-17, Pasadena, CA 91125, USA}

\author[0000-0002-3254-9044]{Karl Glazebrook}
\affiliation{Centre for Astrophysics and Supercomputing, Swinburne University of Technology, Hawthorn, VIC 3122, Australia}
\affiliation{ARC Centre of Excellence for All Sky Astrophysics in 3 Dimensions (ASTRO 3D), Australia}

\author[0000-0002-2784-564X]{Tania M.\ Barone}
\affiliation{Centre for Astrophysics and Supercomputing, Swinburne University of Technology, Hawthorn, VIC 3122, Australia}
\affiliation{ARC Centre of Excellence for All Sky Astrophysics in 3 Dimensions (ASTRO 3D), Australia}

\author[0000-0003-0516-3485]{Hannah Skobe}
\affiliation{Department of Astronomy \& Astrophysics, University of Chicago, Chicago, IL 606374, USA}

\author[0000-0002-1576-2505]{Sarah M.\ Sweet}
\affiliation{School of Mathematics and Physics, University of Queensland, Brisbane, QLD 4072, Australia}
\affiliation{ARC Centre of Excellence for All Sky Astrophysics in 3 Dimensions (ASTRO 3D), Australia}

\author[0000-0003-2804-0648]{Themiya Nanayakkara}
\affiliation{Centre for Astrophysics and Supercomputing, Swinburne University of Technology, Hawthorn, VIC 3122, Australia}

\author[0000-0002-7278-9528]{Rebecca J.\ Allen}
\affiliation{Centre for Astrophysics and Supercomputing, Swinburne University of Technology, Hawthorn, VIC 3122, Australia}

\author[0000-0002-2645-679X]{Keerthi Vasan G.\ C.}
\affiliation{Department of Physics and Astronomy, University of California, Davis, One Shields Ave, Davis, CA 95616, USA}

\author[0000-0001-5860-3419]{Tucker Jones}
\affiliation{Department of Physics and Astronomy, University of California, Davis, One Shields Ave, Davis, CA 95616, USA}

\author[0000-0003-1362-9302]{Glenn G.\ Kacprzak}
\affiliation{Centre for Astrophysics and Supercomputing, Swinburne University of Technology, Hawthorn, VIC 3122, Australia}
\affiliation{ARC Centre of Excellence for All Sky Astrophysics in 3 Dimensions (ASTRO 3D), Australia}

\author[0000-0001-9208-2143]{Kim-Vy H.\ Tran}
\affiliation{School of Physics, University of New South Wales, Kensington, Australia}
\affiliation{ARC Centre of Excellence for All Sky Astrophysics in 3 Dimensions (ASTRO 3D), Australia}

\author{Colin Jacobs}
\affiliation{Centre for Astrophysics and Supercomputing, Swinburne University of Technology, Hawthorn, VIC 3122, Australia}

\begin{abstract}

Gravitational lenses can magnify distant galaxies, allowing us to discover and characterize the stellar populations of intrinsically faint, quiescent galaxies that are otherwise extremely difficult to directly observe at high redshift from ground-based telescopes. Here, we present the spectral analysis of two lensed, quiescent galaxies at $z\gtrsim 1$ discovered by the \texttt{ASTRO 3D Galaxy Evolution with Lenses} survey: \rs\ ($M_*\sim 10^{11.1}M_{\odot}$, $z=1.016$, $\mu \sim 14.6$)  and \reb\ ($M_*\sim 10^{11.5}M_{\odot}$, $z=1.374$, $\mu \sim 4.3$). We measured the age, [Fe/H], and [Mg/Fe] of the two lensed galaxies using deep, rest-frame-optical spectra (S/N $\gtrsim 40$~\AA$^{-1}$) obtained on the Keck~I telescope. The ages of \rs\ and \reb\ are $5.6^{+0.8}_{-0.8}$~Gyr and $3.1^{+0.8}_{-0.3}$~Gyr, respectively, indicating that most of the stars in the galaxies were formed less than 2~Gyr after the Big Bang. Compared to nearby quiescent galaxies of similar masses, the lensed galaxies have lower [Fe/H] and [Mg/H]\@. Surprisingly, the two galaxies have comparable [Mg/Fe] to similar-mass galaxies at lower redshifts, despite their old ages. Using a simple analytic chemical evolution model connecting the instantaneously recycled element Mg with the mass-loading factors of outflows averaged over the entire star formation history, we found that the lensed galaxies may have experienced enhanced outflows during their star formation compared to lower-redshift galaxies, which may explain why they quenched early. 

\end{abstract}


\section{Introduction} \label{sec:intro}
How galaxies form and evolve to the state we see today has been a long-standing question in modern astronomy. One way to unveil the evolutionary history of a galaxy is to constrain its chemical composition because the overall metal abundance of a galaxy reflects the interplay among the gravitational potential \cite[e.g.,][]{Dekel1986}, star formation efficiency \cite[e.g.,][]{Calura2009,Magrini2012} and galactic inflows/outflows \cite[e.g.,][]{Finlator2008}. The well-known mass--metallicity relation (MZR), a tight correlation between galaxy stellar mass and metallicity, has been widely used to investigate the metal retention of galaxies at different masses and redshifts \cite[e.g.,][]{Tremonti2004, Gallazzi05, Erb06, Finlator2008, Kirby2013, Lu15, Ma16, Kacprzak16, Leethochawalit2019, Zhuang2021}.

While the majority of previous studies have used metallicity (i.e., gas-phase oxygen abundance or stellar-phase iron abundance) to characterize the chemical abundances of galaxies that cannot be resolved into stars, spectroscopic studies have turned to detailed abundances of galaxies using modern techniques, such as full-spectrum fitting of the stellar continuum to determine stellar abundances \citep[e.g.,][]{Conroy2018}. Measuring the abundances of individual $\alpha$ elements (including O, Mg, Si and Ca) in stellar populations is a more effective probe of galactic chemical evolution than measuring iron alone. While iron is primarily produced in Type Ia supernovae of low-mass stars with a delayed explosion timescale, $\alpha$-elements are synthesized by core-collapse supernovae of massive stars. Because of the difference in their recycling time, bulk $\alpha$ enhancement---[$\alpha$/Fe]---has been used to measure the duration of star formation historically \citep[e.g.,][]{Thomas2005,Walcher2015, Kriek16}, if assuming a closed box where outflows and/or inflows are absent. In addition, $\alpha$ elements can be approximated to be instantaneously recycled given the short lifetime of massive stars, so they can be used in some simple chemical evolution models. For instance, \citet{Leethochawalit2019} used the relation between stellar masses and [Mg/H] for quiescent galaxies derived from chemical evolution models to constrain the mass-loading factors of galactic outflows, when galaxies are assumed to be leaky boxes in this case. 

Whereas the stellar abundances of quiescent galaxies have been extensively studied out to $z\sim 0.7$ \citep{Choi14,Gallazzi14, Leethochawalit2018,Leethochawalit2019, Beverage21}, we still have limited knowledge about the chemical composition of stellar populations at $z>1$, especially $\alpha$-enhancements. A more nuanced understanding of the chemical abundances of high-$z$ galaxies is crucial to develop a full picture of the chemical evolution and enrichment of galaxies through cosmic time. Unlike measuring the gas-phase abundances for high-$z$ star-forming galaxies, which relies on strong emission lines, constraining the stellar abundances for high-$z$ quiescent galaxies requires deep spectroscopy to capture the faint stellar absorption lines. Consequently, the few available measurements at $z>1$ are either based on a stacked spectrum of quiescent galaxies at similar redshifts and masses \cite[e.g.,][]{Onodera15,Carnall22} or ultra-deep spectroscopy, which typically requires tens of hours on large ground-based 8-10~m telescopes for a single galaxy \cite[e.g.,][]{Toft12, Lonoce15, Kriek16, Kriek19}. Several studies have also attempted to determine the stellar metallicity of high-$z$ quiescent galaxies using low-resolution grism spectra obtained with the \textit{Hubble Space Telescope} (HST) \cite[e.g.,][]{Morishita18,Estrada-Carpenter19}, but these studies primarily used the continuum shape.  Consequently, they cannot characterize the detailed shapes of stellar absorption lines, leading to high degeneracy between stellar metallicity and other galaxy properties, such as age. 

Gravitational lensing is a promising way to revolutionize our understanding of faint, high-$z$ quiescent galaxies because it magnifies both the flux and angular resolution of distant galaxies. The magnified flux of lensed galaxies enabled us to obtain comparable or even higher signal-to-noise (S/N) spectra with shorter integration times, allowing us to investigate the stellar population from the faint stellar continua more accurately \citep[e.g.,][]{Newman18, Jafariyazani20, Man21}. The magnified, lensed galaxies are more extended, providing us an opportunity to resolve the galaxies that are much more compact in the source plane in exquisite detail. With the help of gravitational lensing, early studies were able to reveal unprecedented details of high-$z$ galaxies, such as the metallicity gradient \cite[e.g.,][]{Leethochawalit16, Jafariyazani20} and morphology \cite[e.g.,][]{Newman18, Man21}. 
 
In this work, we present the spectral analysis of two gravitationally lensed, quiescent galaxies at $z\gtrsim 1$ discovered by the \texttt{ASTRO 3D Galaxy Evolution with Lenses} (\agel\@) survey \citep{Tran22} in order to characterize their stellar populations, especially the stellar elemental abundances. In Section \ref{sec:sample_data}, we describe our sample and data used for analysis. We explain how we determine the stellar mass and stellar population properties in Section \ref{sec:methods}. We present our results (the measured stellar population age, [Fe/H] and [Mg/Fe]) in Section \ref{sec:results} and discuss their physical implications in Section \ref{sec:discussion}. Finally, we summarize our findings in Section~\ref{sec:summary}. Throughout this work we assume a flat $\Lambda$CDM cosmology with $\Omega_m=0.3$, $\Omega_{\Lambda}=0.7$ and $H_0=70$~km s$^{-1}$~Mpc$^{-1}$.

\section{Galaxy Sample and Data}\label{sec:sample_data}
\subsection{The Lensed Galaxies}
The two gravitationally lensed galaxies, \agel132304+034319 (hereafter \rs) at $z=1.016$ and \agel001424+004145 (\reb) at $z=1.374$, were identified as part of the \agel\ survey \citep{Tran22}. The \agel\ survey performed spectroscopic observations on strong gravitational lenses selected from the Dark Energy Survey \cite[DES;][]{Abbott18} and the Dark Energy Camera Legacy Survey \cite[DECaLS;][]{Dey19} to measure their redshifts.  These lensed galaxy candidates were discovered using convolutional neural networks \citep{Jacobs19a, Jacobs19b} and were notable for their red arcs. \rs\ was also independently discovered by the COOL-LAMPS Surveys as COOL J1323+0343 \citep{Sukay22} in a visual search of the northern galactic cap portion of the southern DECaLS dataset.

\begin{figure}[t]
    \centering
    \includegraphics[width=0.6\columnwidth]{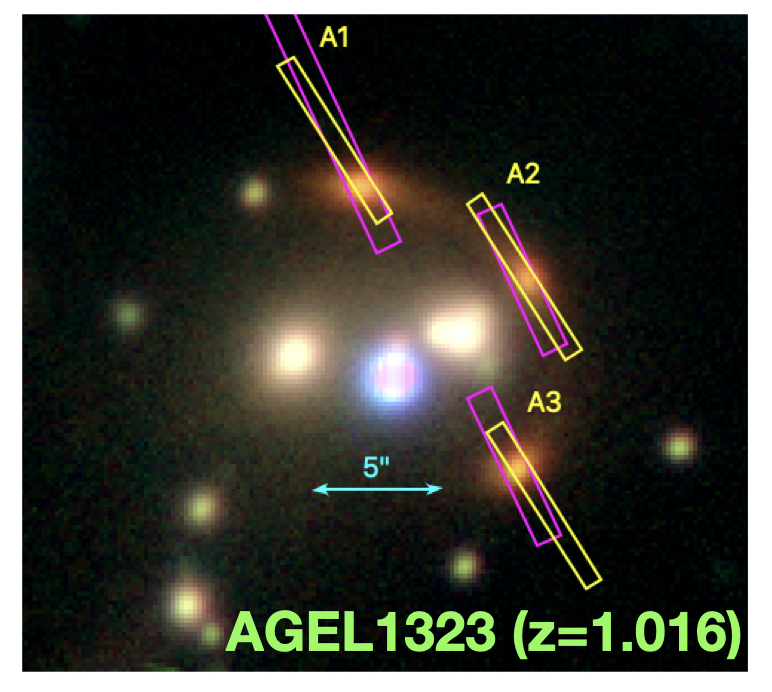}
        \includegraphics[width=0.6\columnwidth]{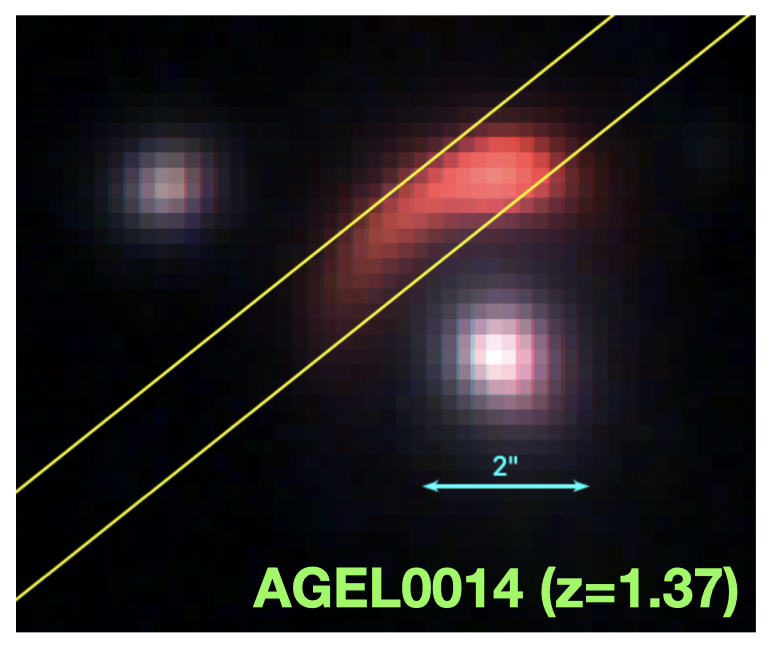}
    \caption{Color images of \rs\ (top) and \reb\ (bottom). The images were constructed using HSC $ri$-band and MOSFIRE-K$_s$ imaging.  The MOSFIRE (yellow) and LRIS (magenta) slits are overplotted. }
    \label{fig:color_img}
\end{figure}

Neither galaxy displayed emission lines in the identification spectra from \agel\ survey, indicating their quiescent nature. As shown in Figure~\ref{fig:color_img}, \rs\ is a five-image\footnote{Two of the lensed images are only visible in high resolution HST imaging (Figure~\ref{fig:RS_HST_color}).} system lensed by a group of galaxies with three visible clumps, while \reb\ is an arc spanning at least 3\arcsec\@ lensed by a foreground lens galaxy. The magnified flux of the lensed images make the two galaxies ideal for studying the detailed abundances and spatially resolved kinematics with deep spectroscopy. While this paper focuses on the chemical abundances of the lensed galaxies, Sweet et al. (2023, in prep.) will present a high-resolution kinematic analysis of \rs\ in an upcoming paper.

In 2021, we acquired deeper follow-up observations with Keck/LRIS and Keck/MOSFIRE for the galaxies to capture the portion of the rest-frame optical between 3600~\AA\@ and 5500~\AA\@, which is sufficient to recover the ages and Fe and Mg abundances of the stellar population \citep{Leethochawalit2018,Leethochawalit2019,Zhuang2021}. In this section, we describe the spectroscopic and photometric data used for the stellar population analyses (Table \ref{tab:observations}). 

\begin{deluxetable*}{cccccccc}[htb!]
\tablecaption{Spectroscopic and Photometric Observations of the Lensed Galaxies\label{tab:observations}}
\tablehead{\colhead{Object} & \colhead{R.A.} & \colhead{Decl.} & \colhead{$z_{\rm spec}$} & \colhead{Instrument} & \colhead{Date} & \colhead{Integration Time} &  \colhead{S/N$^{a}$}
\\ \colhead{} & \colhead{(J2000)} & \colhead{(J2000)} & \colhead{} & \colhead{} & \colhead{} & \colhead{} & \colhead{(\AA$^{-1}$)} }
\startdata
    \rs\@ & $\rm 13^h 23^m 04^s.1$ &  $\rm +03^d 43^m 19^s.4$ & 1.016 &   Keck/LRIS & 2021-04-05 & 6.0~h & 60 \\
    {} & {} & {}  & {} & Keck/MOSFIRE-Y & 2021-04-17 & 5.0~h & 50 \\
    {} & {} & {}  & {} & $^*$Keck/MOSFIRE-Y & 2021-04-17 & 786~s & - \\
    {} & {} & {}  & {} & $^*$Keck/MOSFIRE-K$_s$ & 2021-04-17 & 105~s & - \\
    {} & {} & {}  & {} & $^*$HST/WFC3-F200LP & 2022-06-18 & 600~s & - \\
    {} & {} & {}  & {} & $^*$HST/WFC3-F140W & 2022-06-18 & 598~s & - \\     \hline
    \reb & $\rm 00^h 14^m 24^s.3$ & $\rm +00^d 41^m 45^s.5$ & 1.374 & Keck/MOSFIRE-Y & 2021-08-17 & 1.5~h & 50\\
    {} & {} & {}  & {} & Keck/MOSFIRE-J & 2021-08-17 & 1.5~h & 40 \\
    {} & {} & {}  & {} & $^*$Keck/MOSFIRE-J & 2021-10-30 & 1830~s & - \\
    {} & {} & {}  & {} & $^*$Keck/MOSFIRE-K$_s$ & 2021-10-30 & 1668~s  & - \\
\enddata
\tablenotetext{a}{S/N of the stacked spectra, including all lensed images. The S/N was calculated as median ratio of the noise array to the flux array in the wavelength range of 4000--5500\AA\@.  For this purpose, the noise array was not augmented by $\chi^2$.  That augmentation applies only to the reported uncertainties on stellar population parameters (see Section~\ref{subsec:full_spectrum_fitting}).} 
\tablenotetext{*}{Photometric observations.} 
\end{deluxetable*}

\subsection{Spectroscopy}
\subsubsection{Keck/LRIS}
We observed \rs\ with LRIS on the Keck~I telescope with the D560 dichroic, 400/8500 grating and 1\arcsec\ slit width for the red side for a total of 6~hours. 
The seeing varied between $\sim 0.8$\arcsec\ and $\sim 1.1$\arcsec\ throughout the night. We determined a FWHM of 6.47~\AA\ from the spectrum of the arc lamp, corresponding to 3.21~\AA\ in the rest frame. The spectral energy distribution of \rs\ is such that the blue side of the LRIS spectrum had insufficient flux for a meaningful analysis.

The data were reduced with PypeIt \citep{pypeit:joss,pypeit:zenodo}, a semi-automatic software that performs flat field correction, cosmic ray removal, wavelength calibration, flexure correction via sky lines, heliocentric corrections, sky subtraction and spectral extraction. Because we used HgNeArCdZn lamps, different from the archived arcs in PypeIt, we calibrated the wavelength solution by visually identifying the arc lines instead of using the automatic line detecting algorithms. The extracted 1D spectra of each image of \rs\ (A1, A2, and A3) were flux-calibrated and coadded separately after the main run of PypeIt. The telluric corrections were performed directly on the coadded science spectra, which fits a fifth-order polynomial model representing the continuum of the galaxy and spectral regions sensitive to telluric absorption to a grid of telluric spectral templates. Readers are referred to PypeIt's website\footnote{PypeIt: \url{https://pypeit.readthedocs.io/en/release/index.html}} for details. The coadded spectra for the three lensed images were stacked with inverse variation weighting to maximize the S/N in order to better characterize the stellar population properties.

\subsubsection{Keck/MOSFIRE}
We observed \rs\ with MOSFIRE Y-band on Keck~I telescope MOSFIRE-Y using a slit width of 0.7\arcsec\ for the slit mask for 5~hours. The seeing was around 1\arcsec\@, yielding a FWHM of $\sim 2.5$~\AA\@. 
On the second night, we observed \reb\ using a slit width of 1.0\arcsec\ for a total of 3~hours in Y and J band.  The seeing was around  0.4\arcsec\ on the second night, providing a FHWM of $\sim$ 3.5~\AA\@ in the Y-band and $\sim$ 4.4~\AA\ in the J-band.

The raw data were reduced using the MOSFIRE DRP\footnote{MOSFIRE DRP: \url{https://keck-datareductionpipelines.github.io/MosfireDRP/}} to generate a coadded 2D spectrum for the entire night. Generally, the MOSFIRE DRP expects that the data were taken with a standard dithering pattern of ABAB within the slits, which results in one positive and one negative trace in the differential image. However, we adopted a non-standard dither pattern for \rs\ because the small separations ($\sim 7\arcsec$) between the three lensed images do not allow enough room to nod within the slits. We therefore required the telescope to nod across the slits. The mask was designed to have two pairs of slits separated by 23.94\arcsec, corresponding to the length of three slits. Three traces of the three lensed images were present on the different slits at the two different telescope pointings. This strategy made the differential image consist of either a positive or a negative trace in one slit. We modified the DRP slightly to handle the sky subtraction to account for this nodding solution. In each slit of the differential image, the light profile was generated from the median of 2D spectrum. We then fit a Gaussian profile to the light profile and masked out the regions within 2$\sigma$ where most of the light is from the galaxy. We then estimated the sky background from the unmasked regions. We also experimented with a few choices of thresholds and found that 2$\sigma$ masking could give the cleanest background while keeping enough signal from the galaxy.

One-dimensional (1D) spectra were extracted from the 2D spectrum produced by the DRP using MOSPEC \citep{Strom2017}. We used optimal extraction \citep{Horne86}. The flux calibration was performed by comparing the observed spectrum of the standard star with the Vega spectrum scaled to its J-band magnitude. A B-spline was fitted to the ratio of the scaled spectrum to the observed spectrum to derive the response curve. We then applied the curve to other extracted science spectra to obtain the flux-calibrated, telluric-corrected spectrum of each science target.

\subsection{Photometry}
\subsubsection{Ground-based Imaging}
To construct the lens model, we obtained MOSFIRE images for \rs\ and \reb\ on 2021 Apr 5 and 2021 Oct 30, respectively. The sky was clear on both nights, with an average seeing of 0.82\arcsec\ on Apr 5 and 0.65\arcsec\ on Oct 30. 
We reduced the images by using the standard IRAF commands to perform dark subtraction, flat correction, and sky subtraction. The final images have a $5\sigma$ limiting magnitude of 24.88~mag in the J-band and 24.38~mag in the K$_s$-band for \reb, and 23.93~mag in the Y-band and 23.14~mag in the K$_s$-band for \rs\@ on a pixel-to-pixel basis. We performed the flux calibration by comparing the instrumental magnitudes of 2MASS stars in the field obtained from \texttt{SExtractor} \citep{Bertin96} with the values in 2MASS Point Source Catalog \citep{Skrutskie06}. 

In addition, the public images of \rs\ and \reb\ in the Hyper Suprime-Cam (HSC) Subaru Strategic Program DR3 \citep{Aihara22} were used for lens modeling (Section~\ref{sec:SED_fitting_mass}). 

\subsubsection{HST/WFC3}
\begin{figure}[t]
    \centering
    \includegraphics[width=0.8\columnwidth]{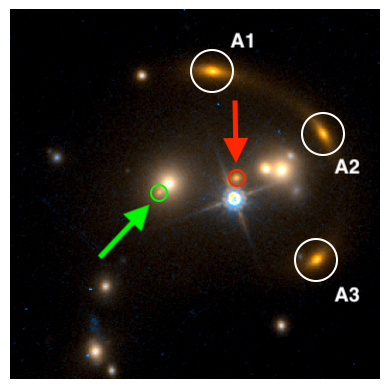}
    \caption{The HST/WFC3 color image of \rs\ constructed with F200LP and F140W imaging. The counter-image and the central image are marked by the green and red circle, respectively.  }
    \label{fig:RS_HST_color}
\end{figure}

We used F140W and F200LP images from HST/WFC3 from program 16773 (PI Glazebrook). The target was observed in the IR/F140W filter for 3 exposures of 200 seconds, and in the UVIS/F200LP filter for 2 exposures of 300 seconds \citep{Shajib22, Shajib22_soft}.  The images were reduced using the STScI DrizzlePac software package to align the separate exposures, as well as correct for background distortion and remove flagged cosmic rays. Within the AstroDrizzle function, we customized the WCS for the final stacked output to rotate the filters in the same orientation. We set the final pixel size of “0.08 to match the F200LP filter with the scaling of filter F140W. To create the mock-RGB combined image in Figure~\ref{fig:RS_HST_color}, we used the \citet{Lupton2004} algorithm implemented in \texttt{astropy}'s \texttt{make\_lupton\_rgb}.

\section{Methods}\label{sec:methods}

\subsection{Lens Modeling}
To obtain a reliable mass estimate, a lens model is needed to calculate the magnification factor and yield the delensed stellar mass. We used the lens modeling software \texttt{PyAutoLens}\footnote{\url{https://github.com/Jammy2211/PyAutoLens}} \citep{Nightingale2015, Nightingale2018, pyautofit, pyautolens} to fit the observed images.

\begin{figure*}
\centering
\includegraphics[width=0.44\textwidth]{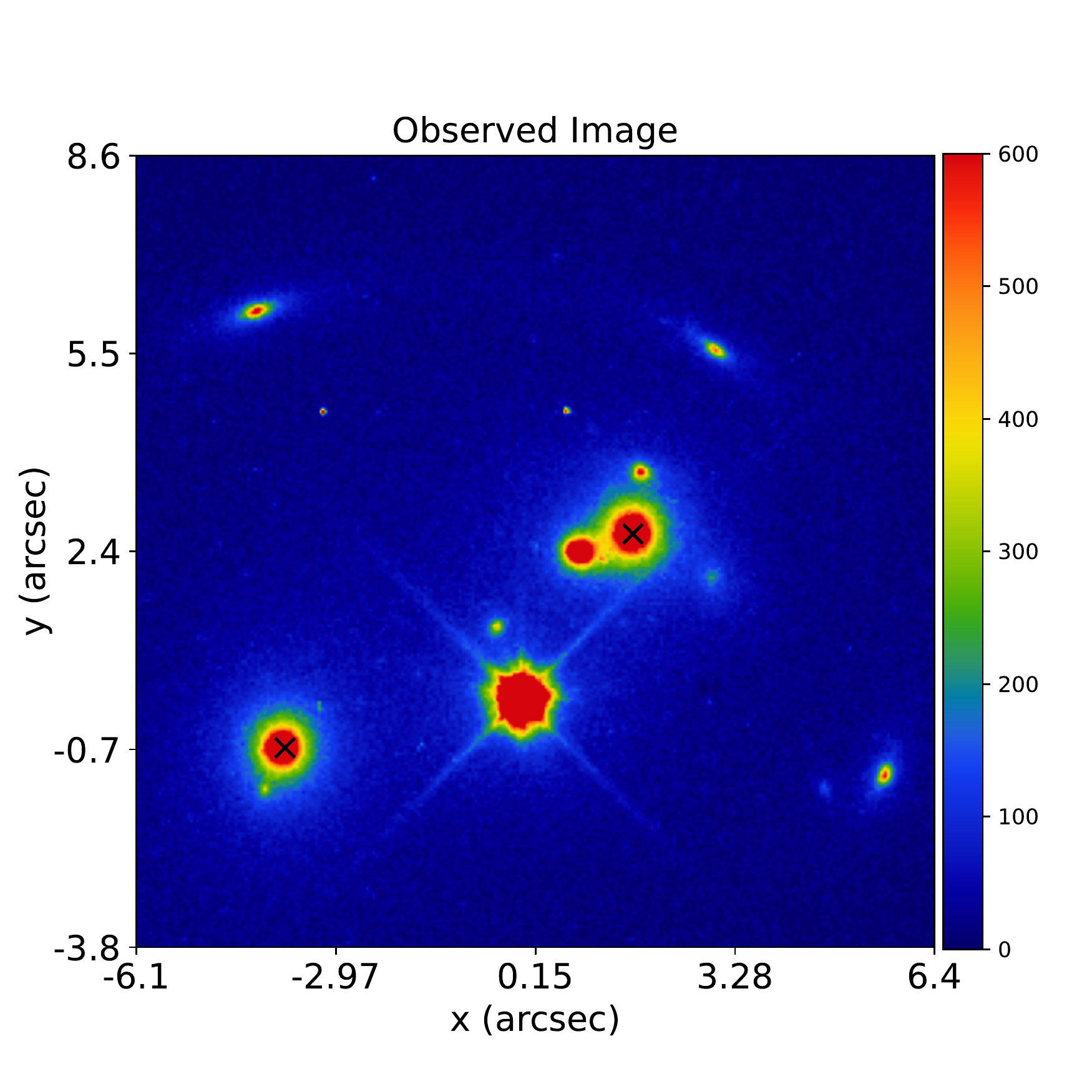}
\includegraphics[width=0.44\textwidth]{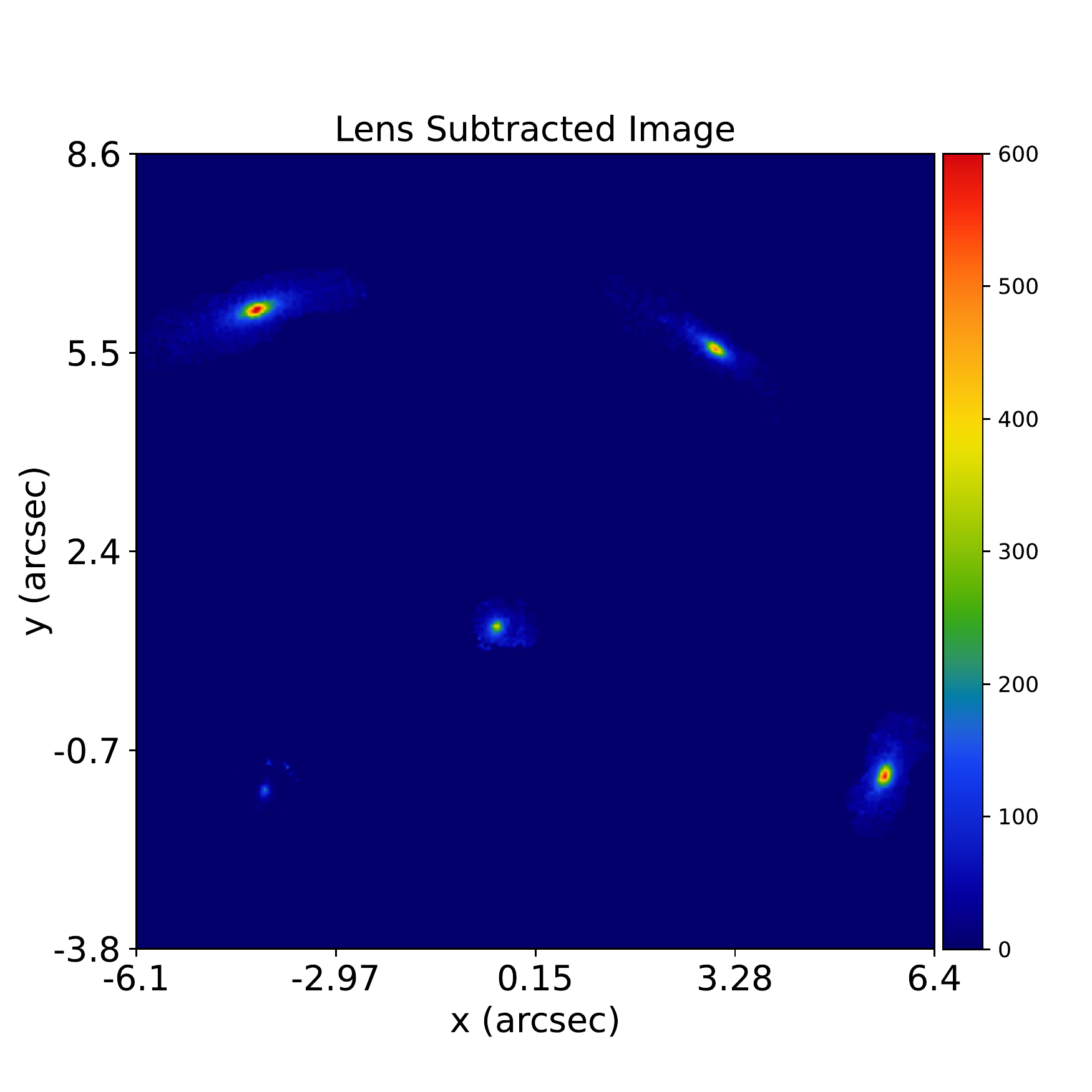}
\includegraphics[width=0.44\textwidth]{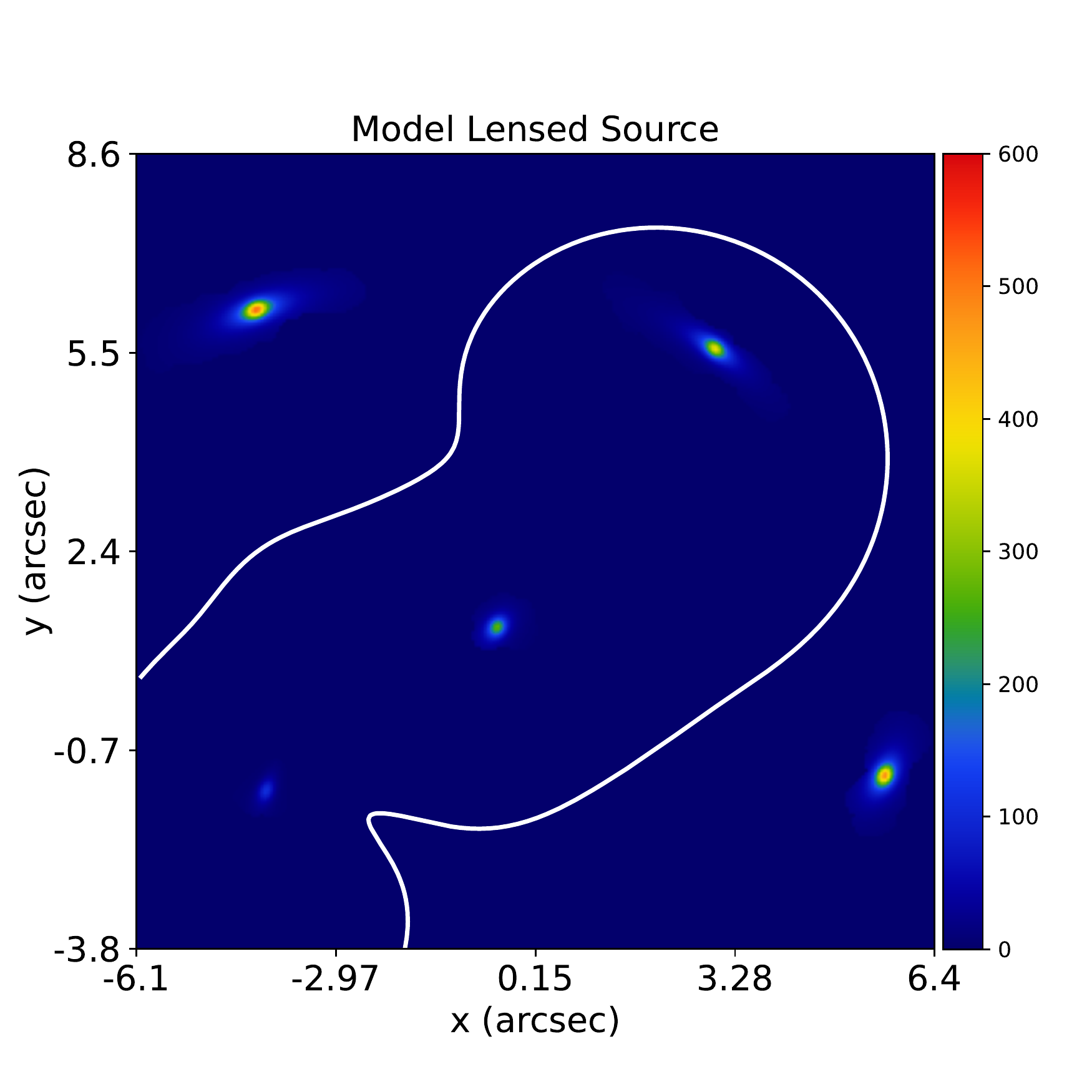}
\includegraphics[width=0.44\textwidth]{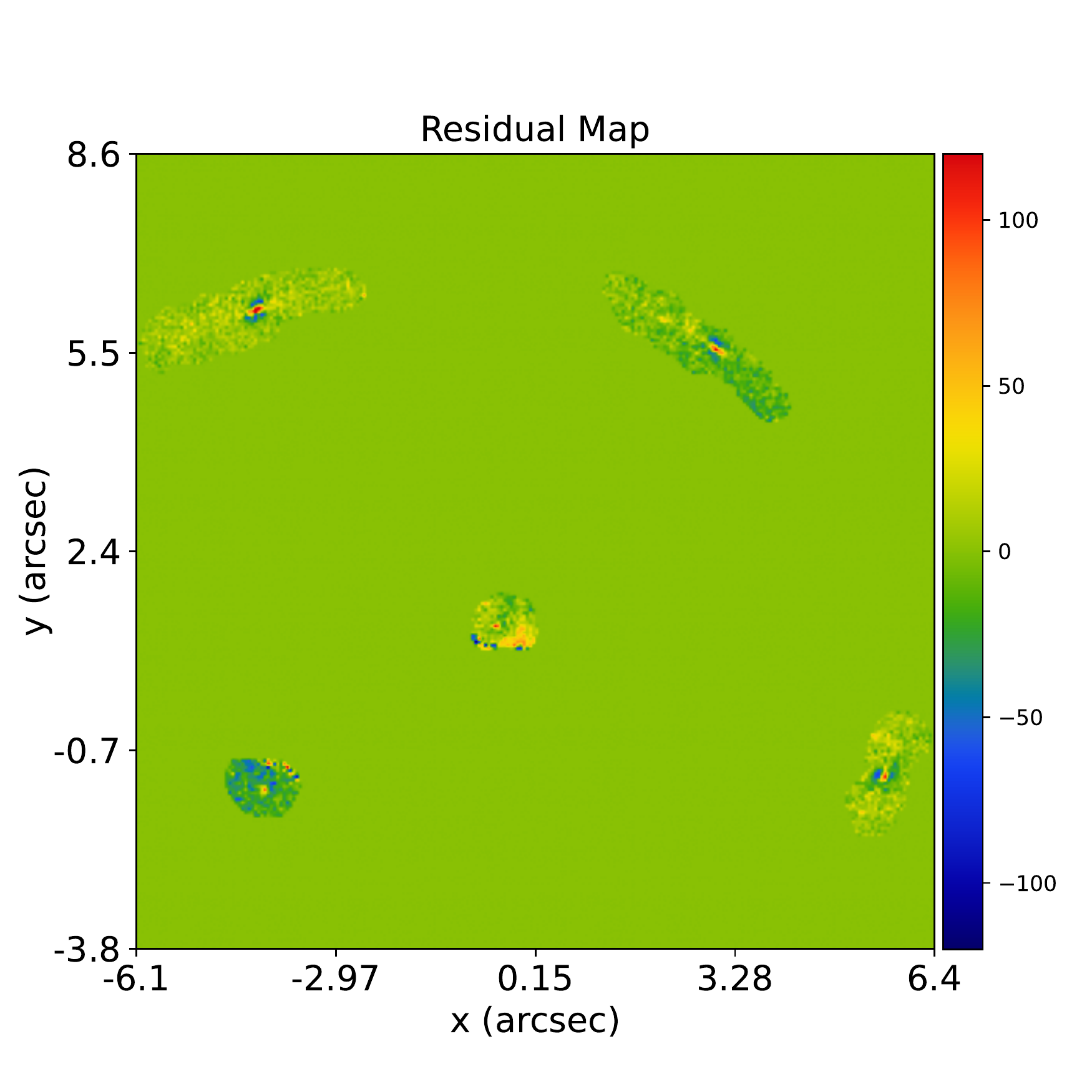}
\caption{
    The observed HST F200 image (upper left), observed image with a model for the foreground lensing galaxies subtracted (upper right), best-fit model lensed source (bottom left) and residual map (bottom right) of lens modeling performed with \texttt{PyAutoLens}. Cut-outs were extracted from the full HST F200 image shown in Figure~\ref{fig:RS_HST_color}. The white line plotted over the bottom left panel is the lens model's tangential critical curve. In the lens-subtracted image, five distinct multiple images are seen, which are all reproduced by the lens model.The units of the colorbar are counts.
    }
\label{fig:rs_lens_model}
\end{figure*}

\begin{figure*}
\centering
\includegraphics[width=0.44\textwidth]{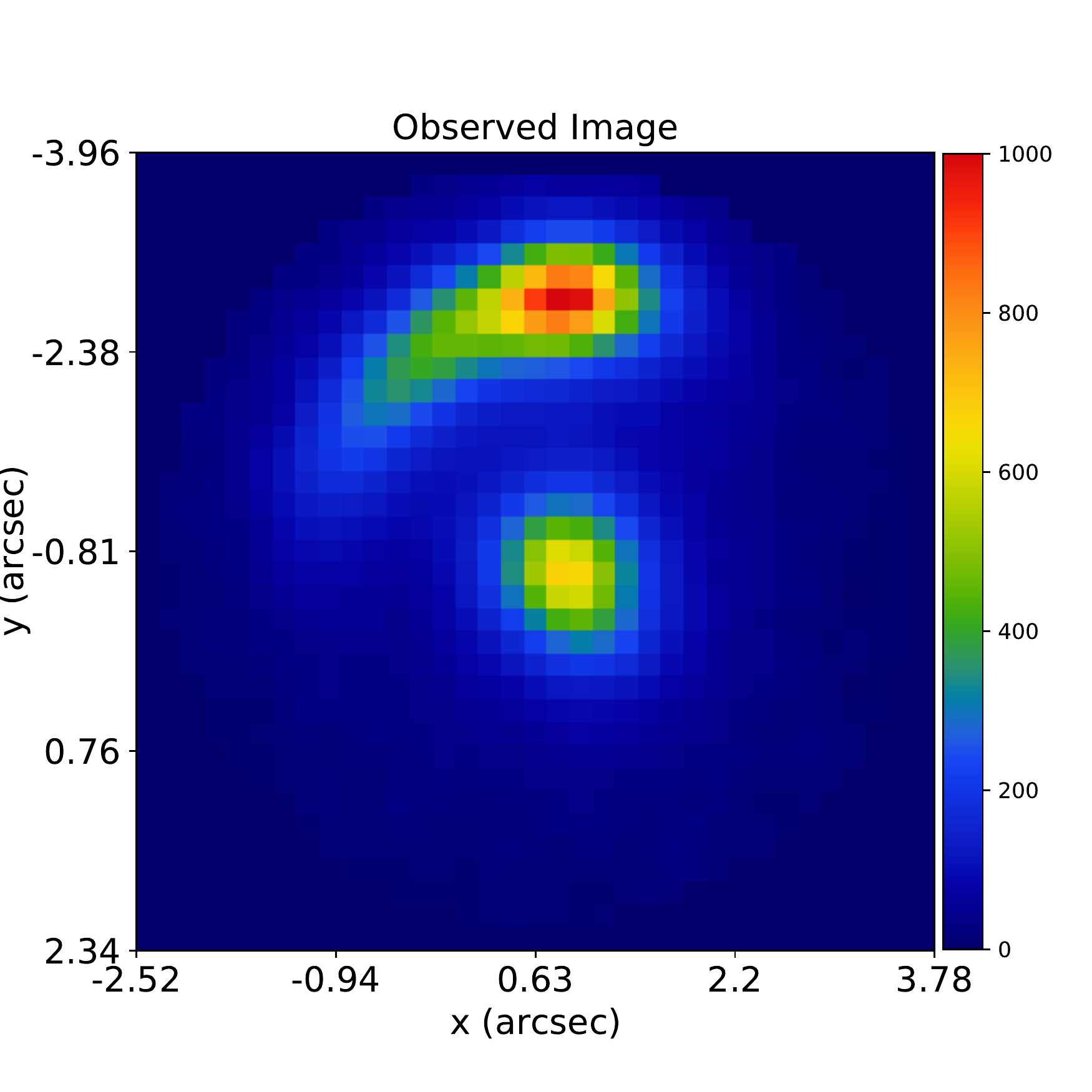}
\includegraphics[width=0.44\textwidth]{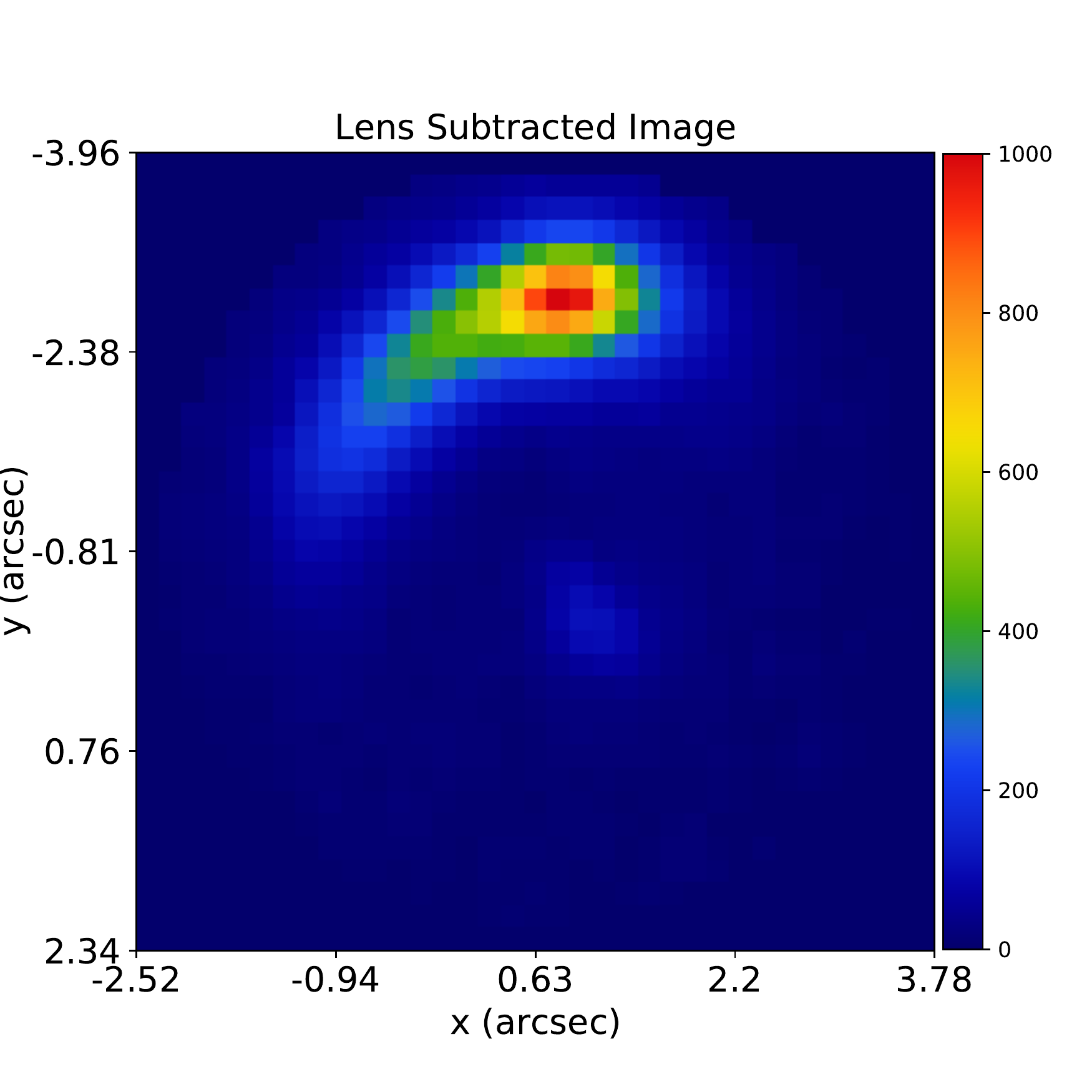}
\includegraphics[width=0.44\textwidth]{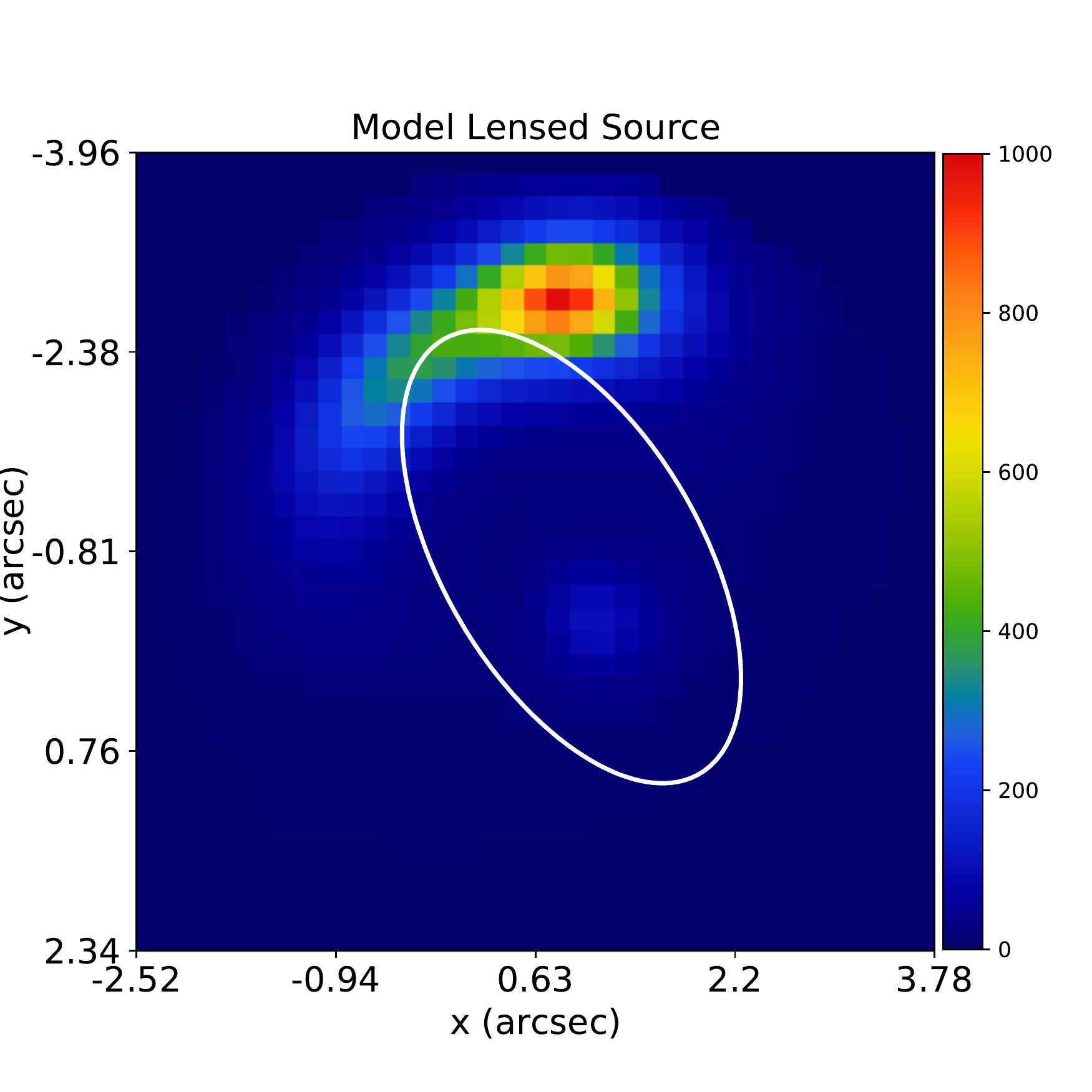}
\includegraphics[width=0.44\textwidth]{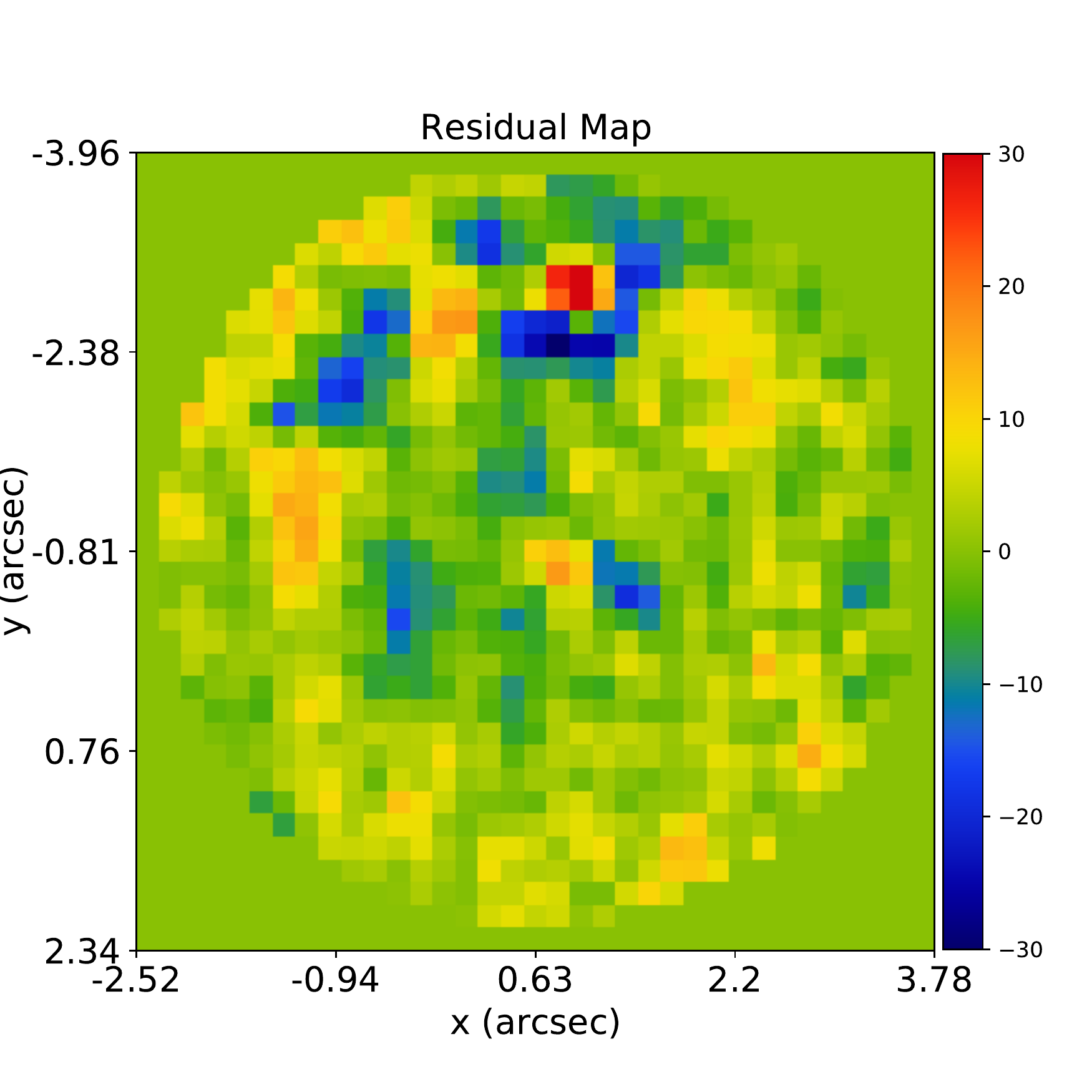}
\caption{
    The observed MOSFIRE-K$_s$ image (upper left), observed image with a model for the foreground lensing galaxy subtracted (upper right), best-fit model lensed source (bottom left) and residual map (bottom right) of lens modeling performed with \texttt{PyAutoLens}. The units of the colorbar are counts. Cut-outs were extracted from the full MOSFIRE-K$_s$ image shown in Figure~\ref{fig:color_img}.  The white line plotted over the bottom left panel is the lens model's tangential critical curve. In the lens-subtracted image, a faint counter-image is seen around $(x=0.7\arcsec, y=-0.7\arcsec)$, which is successfully recovered in the model lensed source.  
    }
\label{fig:reb_lens_model}
\end{figure*}

Although \citet{Sukay22} already constructed the lens model for \rs\ by fitting the Magellan/FourStar $H$-band image with LENSTOOL \citep{Jullo07} and derived a magnification factor of $\mu=74^{+49}_{-28}$, our HST data reveals a fifth image of the source that was not identified by \citet{Sukay22}. The counter-image is very close to one of the lens galaxies and marked by the green arrow in Figure~\ref{fig:RS_HST_color}.  Our results differ from those of \citeauthor{Sukay22}\ because the counter-image is visible only with the high spatial resolution of our HST images.  As a result, our lens model results in a significantly lower magnification.

\rs\ is a group scale lens containing multiple lens galaxies. We fit the HST F200LP imaging data. We first subtract the light of the two brightest galaxies marked with black crosses in Figure~\ref{fig:rs_lens_model}, by fitting elliptical S{\' e}rsic light profiles. To this foreground subtracted data we apply a custom mask (drawn via a graphical user interface) which retains only the lensed source's multiple images. We then fit a mass model where the mass of the two galaxies marked with black crosses in Figure~\ref{fig:rs_lens_model} are singular isothermal ellipsoids, and the group’s host dark matter halo is modeled with a spherical NFW profile \citep{nfw}. The NFW profile’s concentration is set to the mean of the mass-concentration relation of \citet{Ludlow2016}. The source is again modeled by an elliptical S{\' e}rsic profile. We fit the mass and source profiles simultaneously. As shown in Figure~\ref{fig:rs_lens_model}, the best-fit model successfully reproduces all five images, including the central image. We define the magnification $\mu$ as the ratio of the total image-plane flux of the source divided by the total source-plane flux, and infer a value of $\mu = 14.55 ^{+0.55}_{-1.07}$.

For \reb\@, we performed the fitting on the MOSFIRE-K$_s$ image. For the lens galaxy, we fit an elliptical S{\' e}rsic light profile for its light and singular isothermal ellipsoid for its mass. For the source, we fit an elliptical S{\' e}rsic light profile. We fit all three components simultaneously, using the nested sampling algorithm \texttt{dynesty} \citep{dynesty}. As shown in Figure~\ref{fig:reb_lens_model}, the best-fit model successfully detects a faint source counter image, which is offset $\sim 0.5"$ from the lens galaxy center and is not visible until after the lens light subtraction. For the magnification we infer a value of $\mu = 4.33 ^{+0.16}_{-0.11}$.

\subsection{Stellar Mass Estimates}
\label{sec:SED_fitting_mass}

The lensed stellar masses were measured by fitting  the broadband spectral energy distribution (SED)\@. For \rs\, we used photometry from DECaLS $grz$, HSC-$ri$, MOSFIRE-YK$_s$ and HST/WFC3-F140W\@. 
We adopted HSC-$grizy$ and MOSFIRE-YK$_s$ for the photometry of \reb\@. For each galaxy, we smoothed the high resolution images to match the point spread function (PSF) with the one that has lowest resolution and extracted the photometry from the reduced images with \texttt{SExtractor} \citep{Bertin96}. We padded the photometric uncertainties to 0.1~mag if the formal errors are smaller than that to account for possible systematics originated from complicated morphologies of the lensing system. The photometry is corrected for Galactic reddening using the $E(B-V)$ values measured by \citet{Schlafly2011}.

\begin{figure*}[t]
    \centering
    \includegraphics[width=0.55\textwidth]{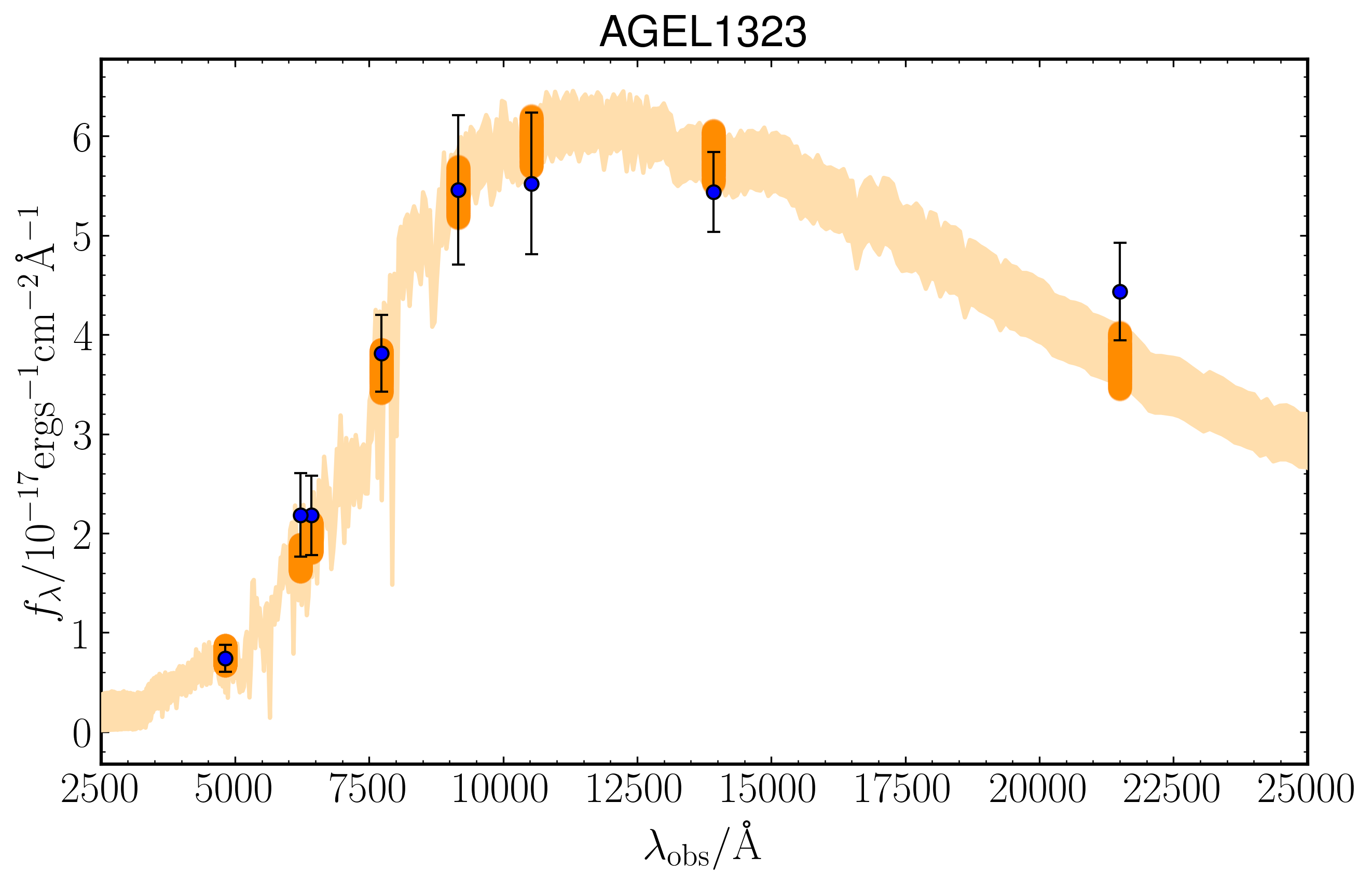}
    \includegraphics[width=0.33\textwidth]{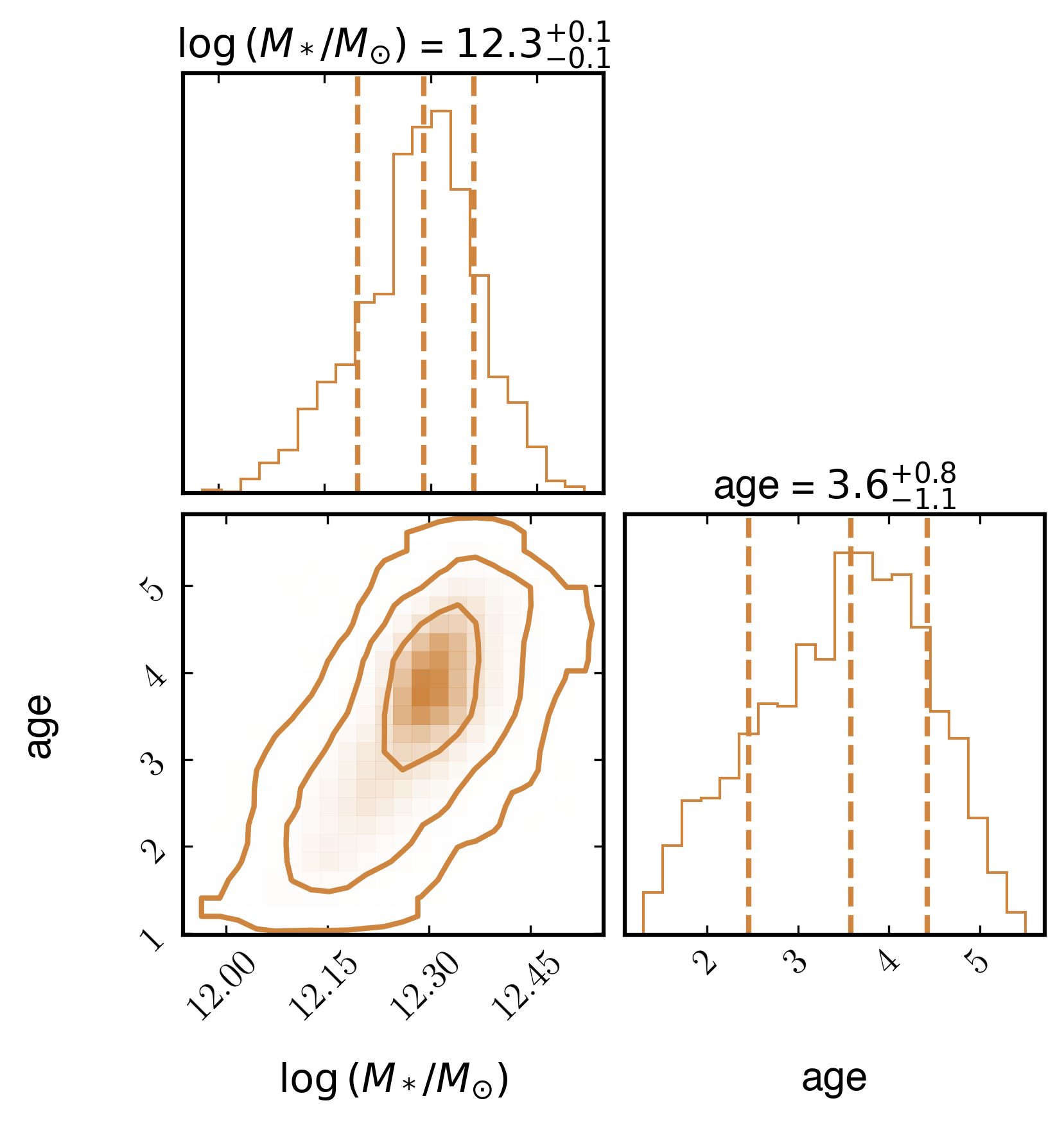}
    \includegraphics[width=0.55\textwidth]{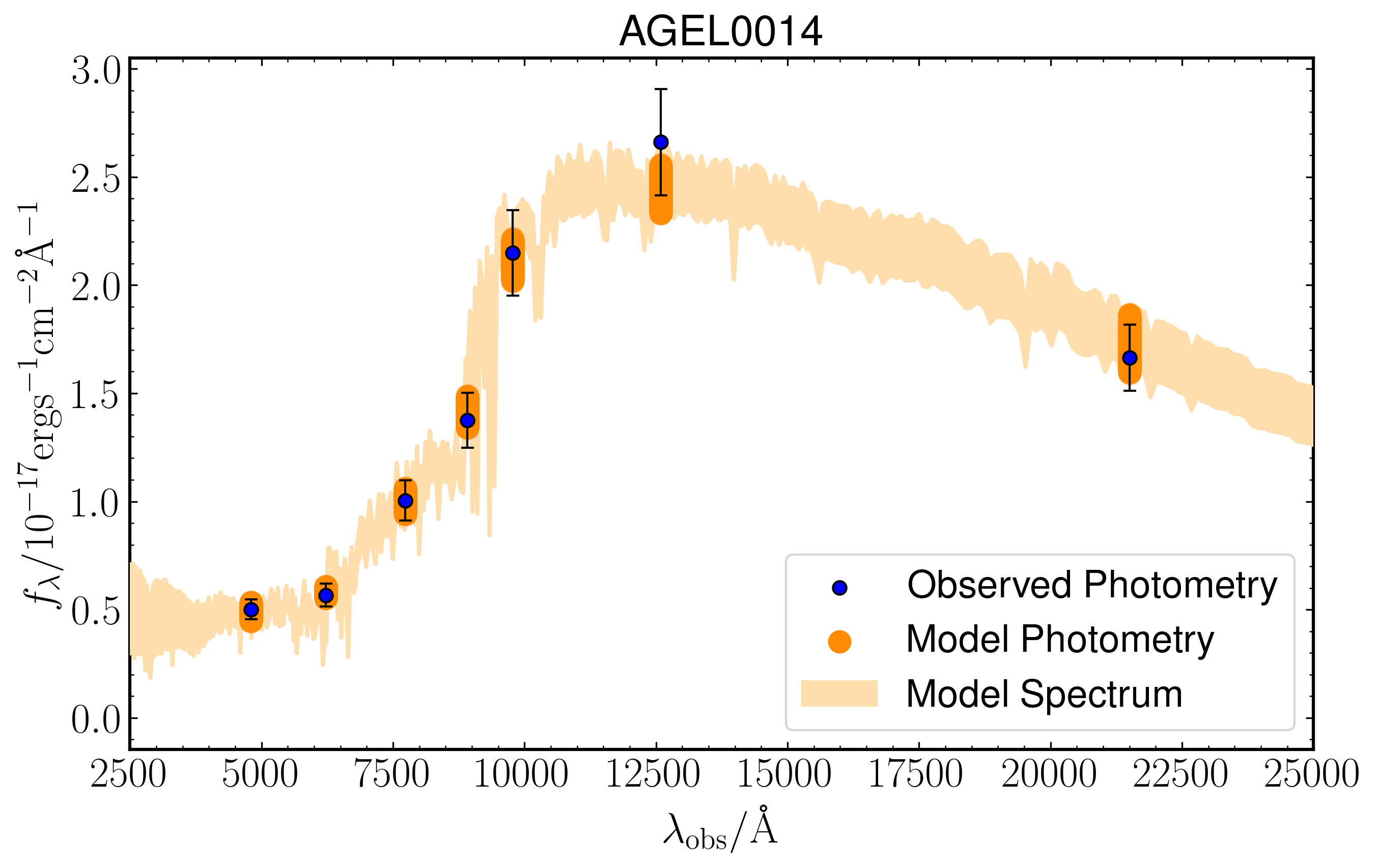}
    \includegraphics[width=0.33\textwidth]{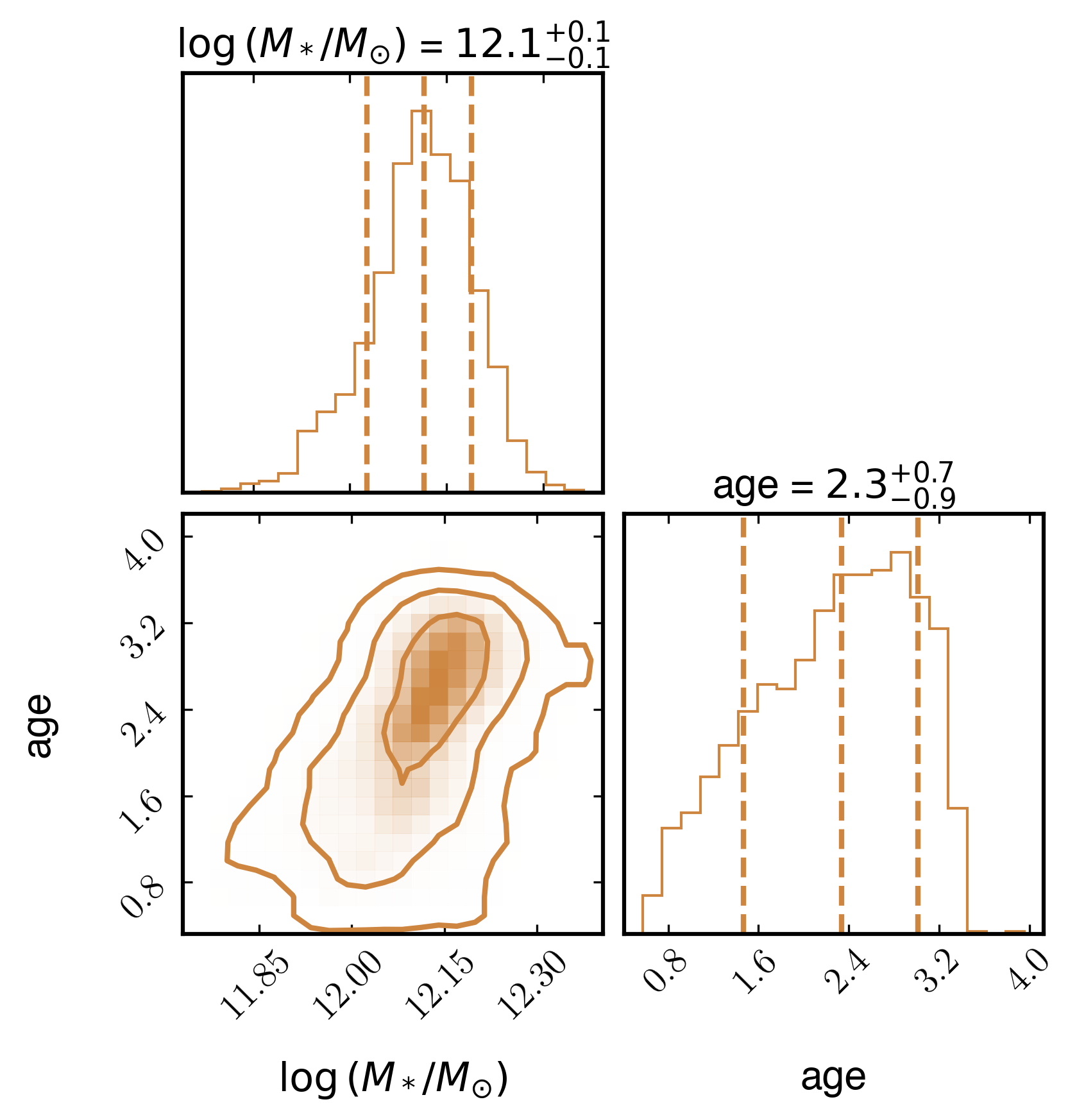}
    \caption{Left: The observed photometry and the SED model of \rs\ (top) and \reb\ (bottom) assuming delayed-$\tau$ SFH derived by BAGPIPES \citep{Carnall18, Carnall19}. The observed and model photometry  are shown by the navy and orange dots, respectively. For \rs\@, the photometric data is from the DECaLS $grz$-band, HSC $ri$-band, MOSFIRE-YK$_s$ and HST/WFC3-F140W. For \reb\@, the photometry is from HSC $grizy$-band and MOSFIRE-JK$_s$. The orange regions show the $1\sigma$ confidence interval of model spectra (light orange) and model photometry (dark orange). Right: The posterior distributions and covariances of the lensed stellar mass and the mass-weighted stellar age averaged over the SFH determined from BAGPIPES for \rs\ (top) and \reb\ (bottom). The contours represent the $1\sigma$, $2\sigma$ and $3\sigma$ level.}
    \label{fig:SED_BAGPIPES}
\end{figure*}

We modeled the SED assuming an delayed exponentially declining (delayed-$\tau$) SFH with BAGPIPES \citep{Carnall18, Carnall19}. We used the \citet{Calzetti2000} dust extinction law and required the attenuation in the $V$ band ($A_V$) to be varied between $0 < A_V < 2$. The redshift for each galaxy has a Gaussian prior centered at $z_{spec}$ in the \agel\ catalog, with a standard deviation of $0.005$. The other free parameters and their priors are stellar mass ($9 < \log{(M_*/M_{\odot})} < 15$), stellar metallicity ($-2 < \log{(Z/Z_{\odot})} < 0.4$), time since the onset of star formation ($30$~Myr $ < T_0 < t_{obs}$\footnote{$t_{obs}$ stands for the age of the universe at the observed redshift.}), and the $e$-folding SFR timescale ($30~\rm Myr < \tau < 10~Gyr$). As shown in Figure~\ref{fig:SED_BAGPIPES}, the SED models yield a lensed stellar mass of $\log{(M_*/M_{\odot})} = 12.3\pm 0.1$ and $\log{(M_*/M_{\odot})} = 12.1\pm 0.1$ for \rs\ and \reb\@, respectively, which translates a delensed stellar mass of $\log{(M_*/M_{\odot})} = 11.13\pm 0.10$ and $\log{(M_*/M_{\odot})} = 11.48\pm 0.09$ when the magnification is corrected.

To quantify the systematic uncertainties of stellar masses due to different assumptions on SFH, we also performed SED fitting with four other SFH models, including single burst, single exponentially declining, log-normal, and double-power-law SFHs. The root-mean-square (RMS) errors among the best-fit stellar masses\footnote{The best-fit stellar masses here mean the median value of the 1D posterior distribution.} of different models are used as estimates of the systematic uncertainties ($0.02$~dex for \rs\ and $0.08$~dex for \reb\@). In the end, we obtained a delensed stellar mass of $\log{(M_*/M_{\odot})} = 11.13\pm 0.10$ for \rs\ and $\log{(M_*/M_{\odot})} = 11.48\pm 0.12$ for \reb\@. We also performed some robustness tests: experimenting with different detection and analysis thresholds for \texttt{SExtractor}, including spectroscopic data for SED fitting, and only fitting for the photometry reported by \citet{Sukay22}. All the results are consistent with our reported values within $1\sigma$.

For both galaxies, the delensed stellar masses of the two galaxies derived from the SED fitting and the velocity dispersion determined from the full-spectrum fitting (Section~\ref{subsec:full_spectrum_fitting})
are consistent with the stellar mass--velocity dispersion relation of quiescent galaxies at $z\sim 1$ \citep{Belli14, Mendel2020}.

\subsection{Stellar Population Fitting}\label{subsec:full_spectrum_fitting}
We analyzed the combined spectra using the full-spectrum fitting algorithm absorption line fitter \cite[\alf\@,][]{Conroy12,Conroy2018} because of its capability of measuring detailed elemental abundances in addition to overall stellar metallicity, [Fe/H]\@. \alf\ uses MIST isochrones \citep{Choi2016} and empirical stellar spectra \citep{Sanchez-Blazquez06,Villaume17}, along with a theoretical response function covering a wide range of elemental abundances. This code operates on the continuum-normalized spectrum by fitting a high-order polynomial to the ratio of the model and data to avoid potential issues with imperfect flux calibration and dust attenuation. This functionality also allows \alf\ to fit multiple spectra of the same source taken on various instruments with different flux calibrations and instrumental resolutions. The fitting is accomplished with the ensemble Markov Chain Monte Carlo (MCMC) sampler \texttt{emcee} \citep{Foreman-Mackey2013} to efficiently explore a large parameter space.   
 
We used \alf's simple mode, which assumes a simple stellar population (SSP) and a fixed \citet{Kroupa2001} initial mass function (IMF)\@. It simultaneously fits for the recession velocity, velocity dispersion, stellar age, stellar metallicity [Fe/H], and the abundances of C, N, O, Na, Mg, Si, Ca, and Ti. The resulting stellar population parameters are SSP-equivalent, which are similar to light-weighted age and abundances. Following \citet{Beverage21}, the default priors were used, except that of maximum age was set to be 1~Gyr older than the age of universe at the observed redshift to fully explore the parameter space and to avoid a truncated posterior distribution. The recovered parameters are consistent with the results when the maximum allowed age is the age of universe at the observed redshift. Although \alf\ also has a full mode describing a double-component stellar population, \citet{Conroy2018} indicated that the simple mode is more reliable when only the blue optical spectrum is available. \citet{Zhuang2021} also found out that the abundances derived from the \alf\ simple mode are more consistent with those obtained from resolved stellar spectroscopy. Therefore, the following analysis focuses on the results obtained from the simple mode.

Given the quality and wavelength coverage of our data, we focused on the age, [Fe/H], and [Mg/Fe] derived from \alf\@. We chose to measure Mg for two reasons. First, Mg is one of the $\alpha$-elements that is mainly produced by core-collapse supernovae. Due to the short lifetime of massive stars, its recycling time can be approximated as instantaneous, for which magnesium is a good tracer of the overall galaxy evolution. For instance, \citet{Leethochawalit2019} used [Mg/H] of quiescent galaxies as a proxy to constrain the mass-loading factor of outflows averaged over the entire SFH\@. Second, the Mg~$b$ triplet at 5170~\AA, which minimally overlaps with absorption features of other elements, is available in our spectra, allowing us to determine the Mg abundance reliably.   

We excluded any regions below 4000~\AA\ for the \alf\ fitting because the difficulty of matching the continuum around the 4000~\AA\ break may introduce some errors in the recovered abundances \footnote{private communication with Meng Gu}. This also excluded the Ca$\,${\sc ii} H and K lines from the fitting, which is known to have strong non-local thermodynamic equilibrium (NLTE) effects that the current models poorly reproduce \citep{Conroy2018}. For each spectrum, we ran two iterations to remove the sky line residuals. In the first iteration, we only masked out the regions where the telluric features cannot be corrected.  After the first fit, the pixels with 4$\sigma$ deviations from the best-fit model were masked to reduce the contamination of bright sky lines to the fitting. We then re-fit the spectrum with the new pixel mask to determine the best-fit model and the corresponding stellar population parameters. The reduced $\chi^2$ of each spectrum is greater than 1, which may result from either models imperfections or underestimated flux uncertainties from the data reduction pipeline. Therefore, the uncertainties reported in Table~\ref{tab:alf_results} are rescaled by the square root of the reduced $\chi^2$ to incorporate the systematic errors that were not captured by the posterior distribution and the random errors. We experimented with different degrees of polynomials for continuum normalization and found out that all the measurements are consistent with each other.

\begin{figure*}[t]
    \centering
    \includegraphics[width=0.56\textwidth]{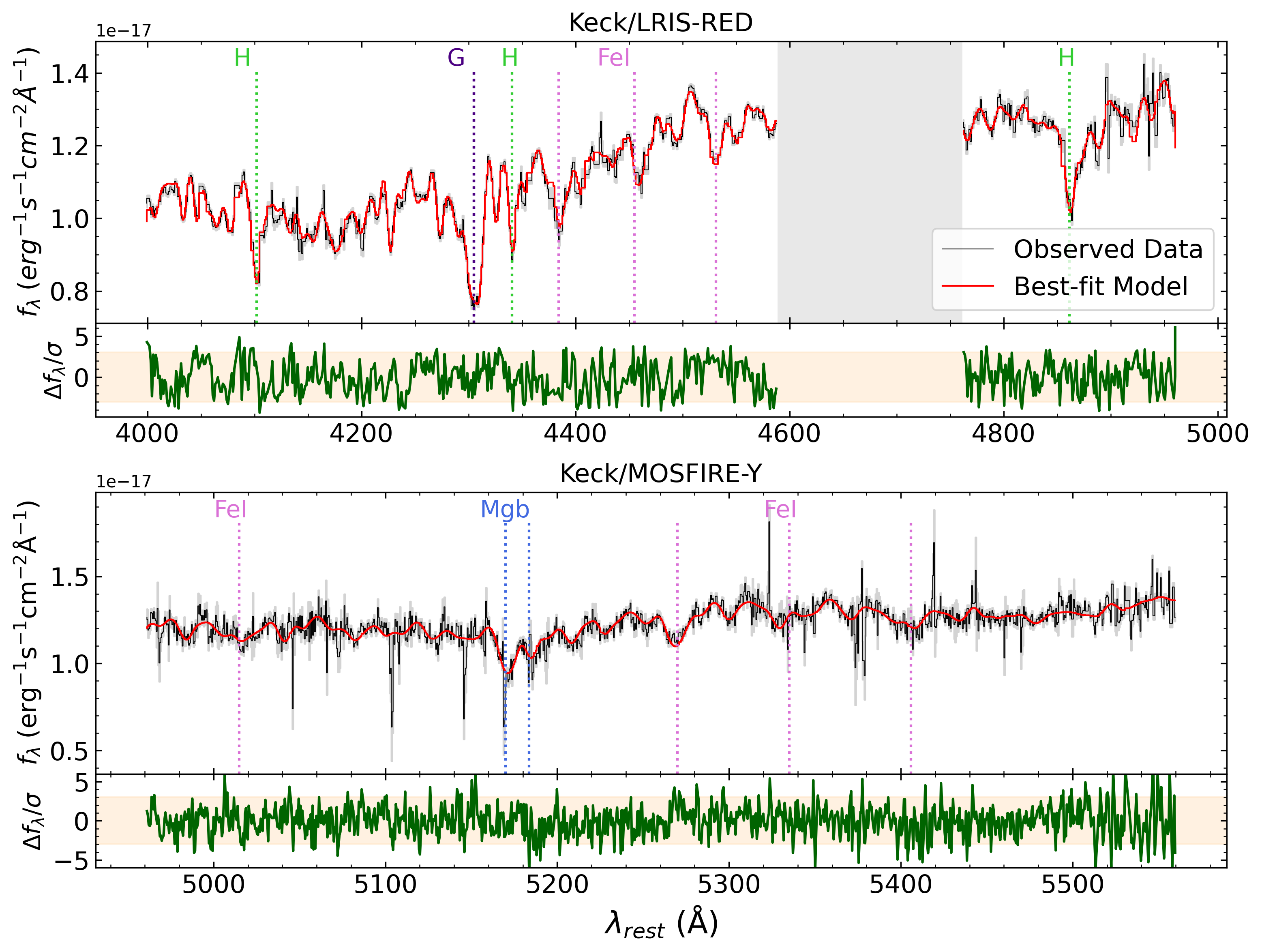}%
    \includegraphics[width=0.42\textwidth]{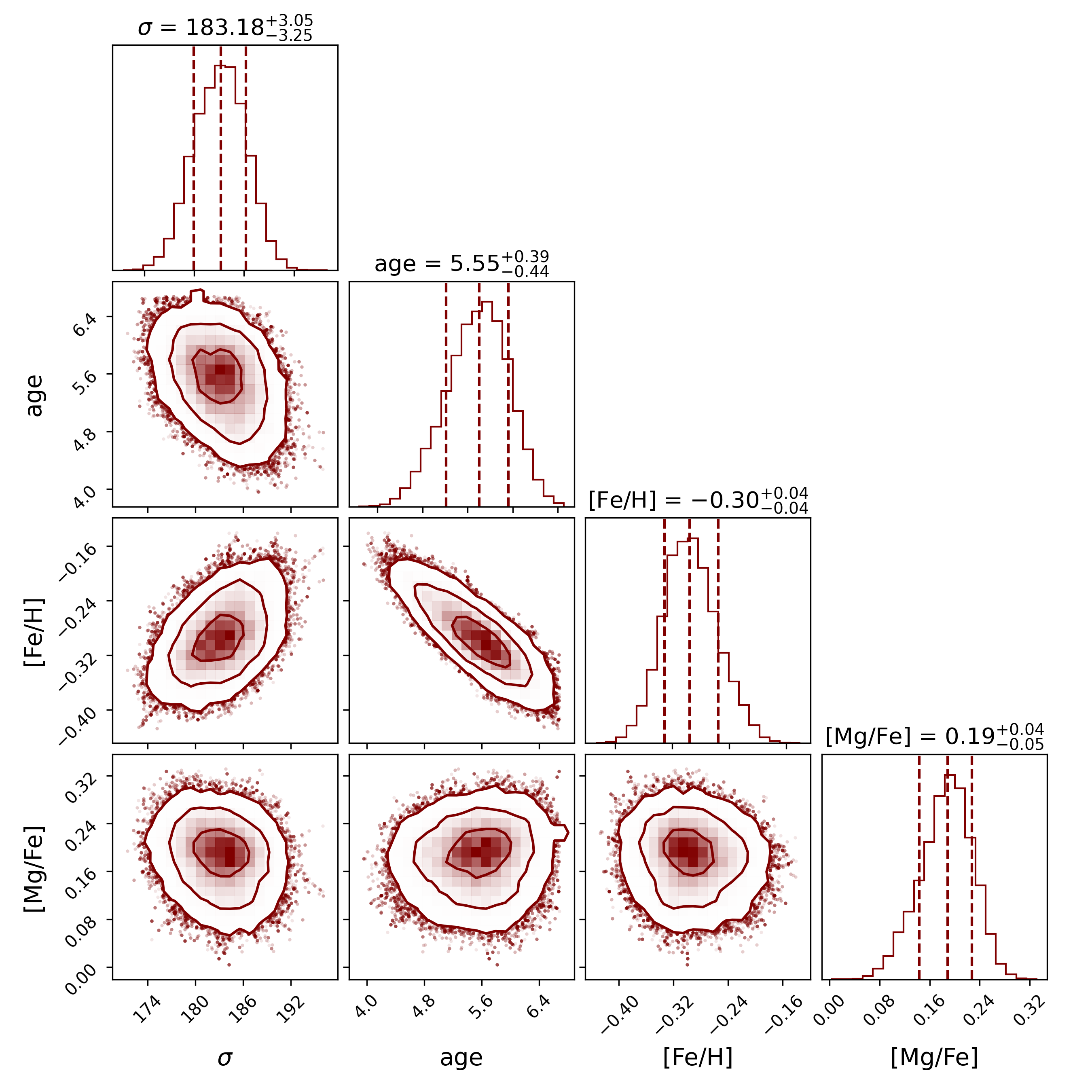}
    \caption{Left: The observed spectrum (black) and best-fit SSP model (red) used in this work for \rs\@. The gray areas represent the $1\sigma$ uncertainties of the observed spectrum. We masked out a small portion in the LRIS-red spectrum for \rs\ because of the telluric features that cannot be corrected. The green lines show the ratio of the model residuals over the uncertainties, with the $3\sigma$ regions indicated by the shaded beige regions. Right: Corner plot showing the posterior distributions and covariances among velocity dispersion ($\sigma$), age, [Fe/H], and [Mg/Fe] of the best-fit model. The uncertainties shown here are the random errors obtained from the posterior distribution, without being corrected for the systematic errors (See Section~\ref{subsec:full_spectrum_fitting}). The contours demonstrate the $1\sigma$, $2\sigma$ and $3\sigma$ levels. }
    \label{fig:spec_alf_posteior_rs}
\end{figure*}

\begin{figure*}[t]
    \centering
   \includegraphics[width=0.56\textwidth]{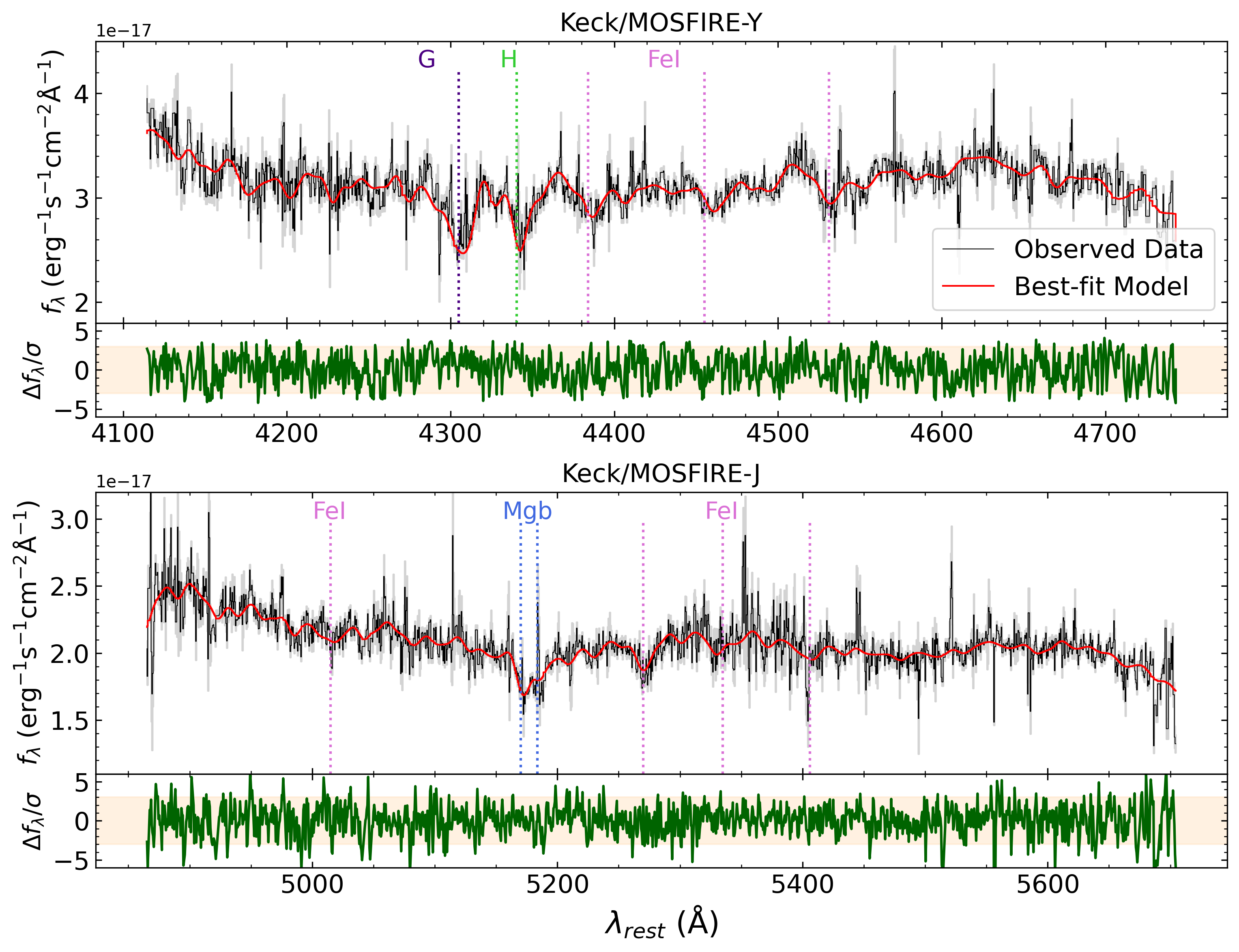}%
        \includegraphics[width=0.42\textwidth]{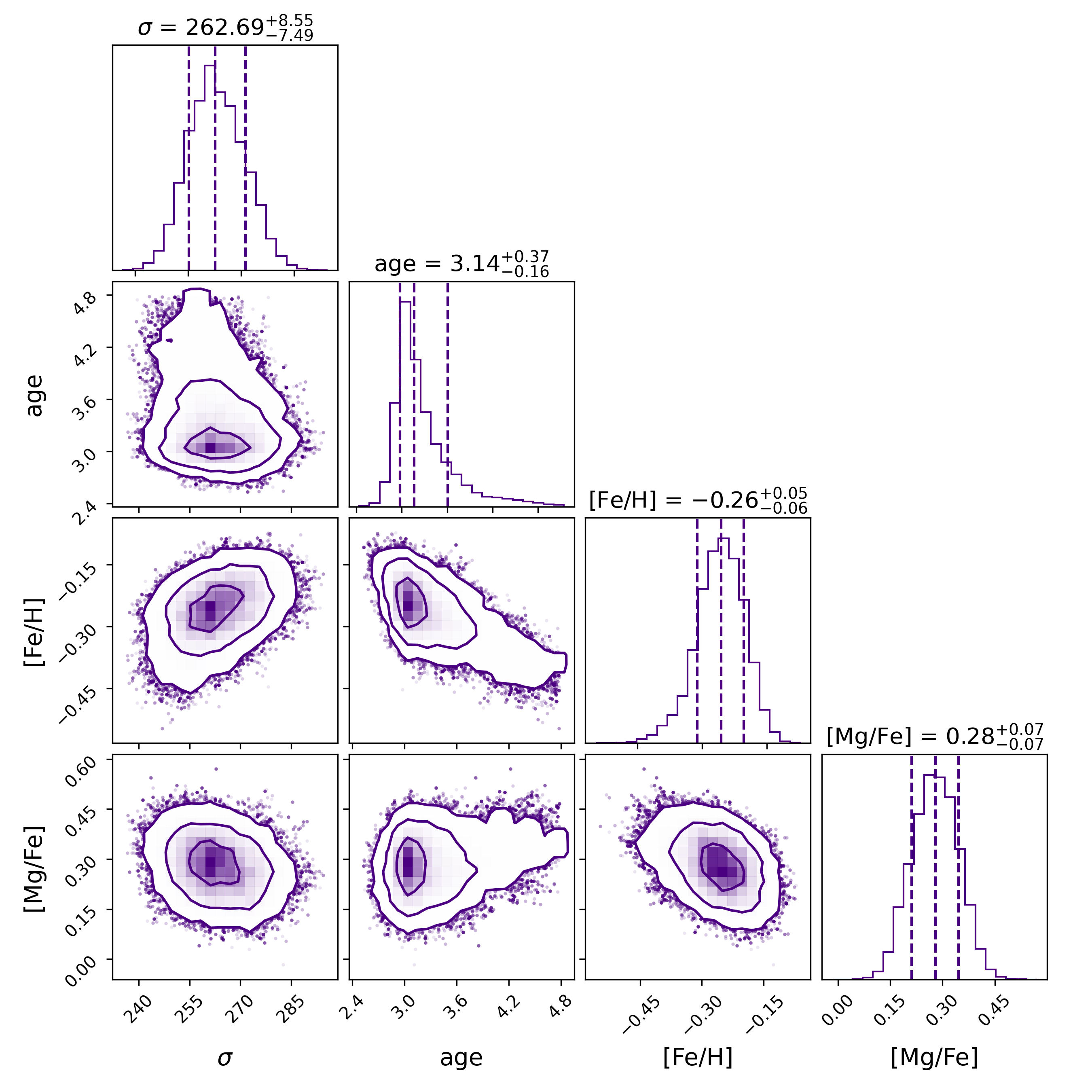}
    \caption{Same as Figure~\ref{fig:spec_alf_posteior_rs}, but showing the results of \reb\@.}
    \label{fig:spec_alf_posteior_reb}
\end{figure*}

The observed spectra and best-fit models determined by \alf's simple mode, as well as the posterior distributions of the corresponding model parameters without the error correction, are shown in Figure~\ref{fig:spec_alf_posteior_rs} for \rs\ and Figure~\ref{fig:spec_alf_posteior_reb} for \reb\@. We list all measured parameters in Table~\ref{tab:alf_results}.

\begin{deluxetable}{ccccc}[t]
\tablecaption{Stellar Population Properties of the Lensed Galaxies\label{tab:alf_results}}
\tablehead{ \colhead{Row} & \colhead{Quantity} & \colhead{Unit}  & \colhead{\rs} & \colhead{\reb} }
\startdata
(1) & $\log{M_*}$ & $M_{\odot}$ & $11.13\pm 0.10$ & $11.48\pm 0.12$\\ \hline
(2) & $\sigma_v$ & km s$^{-1}$ & $183^{+7}_{-8}$ & $263^{+18}_{-16}$\\
(3) & $t_{\rm SSP}$ & Gyr & $5.6^{+0.8}_{-0.8}$ & $3.1^{+0.8}_{-0.3}$ \\
(4) & [Fe/H]$_{\rm SSP}$ & dex & $-0.30^{+0.08}_{-0.07}$ & $-0.26^{+0.10}_{-0.11}$ \\
(5) & [Mg/Fe]$_{\rm SSP}$ & dex & $0.19^{+0.08}_{-0.09}$ & $0.28^{+0.13}_{-0.13}$ \\
(6) & [Mg/H]$_{\rm SSP}$ & dex & $-0.11^{+0.10}_{-0.11}$ & $0.02^{+0.13}_{-0.14}$ \\ 
(7) & $\rm \chi^2_{SSP}/d.o.f$ & - & 3.88 & 3.80 \\\hline
\enddata
\tablecomments{Rows show (1) the delensed stellar mass estimates; (2-7) the velocity dispersion, stellar ages, [Fe/H], [Mg/Fe], [Mg/H] and the reduced $\chi^2$ determined from \alf's simple mode assuming an SSP. The errors here have been corrected for the systematic uncertainties originated from the imperfect models and underestimated noise from the pipelines (See Section~\ref{subsec:full_spectrum_fitting}).}
\end{deluxetable}

\section{Results}\label{sec:results}
\subsection{Old Ages and Possible Rapid Star Formation}

The measured SSP ages of \rs\ and \reb\ are $5.6\pm 0.3$~Gyr and $3.1^{+0.8}_{-0.3}$~Gyr, respectively, both corresponding to a formation redshift of $z_{\rm form}\gtrsim 4$. Although the SSP assumption may be too simple compared to the true SFHs,  the SSP age is still a good indicator of the time when most of stars were formed in a galaxy considering its bias towards the youngest population. In other words, our \alf\ results suggest that the majority of the stellar populations in the two galaxies were formed less than 2~Gyr after the Big Bang, indicating that they must have experienced rapid star formation to reach the stellar masses we see today.

\subsection{Chemical Abundances of the Lensed Galaxies}

\begin{figure}[t]
    \centering
    \includegraphics[width=0.4\textwidth]{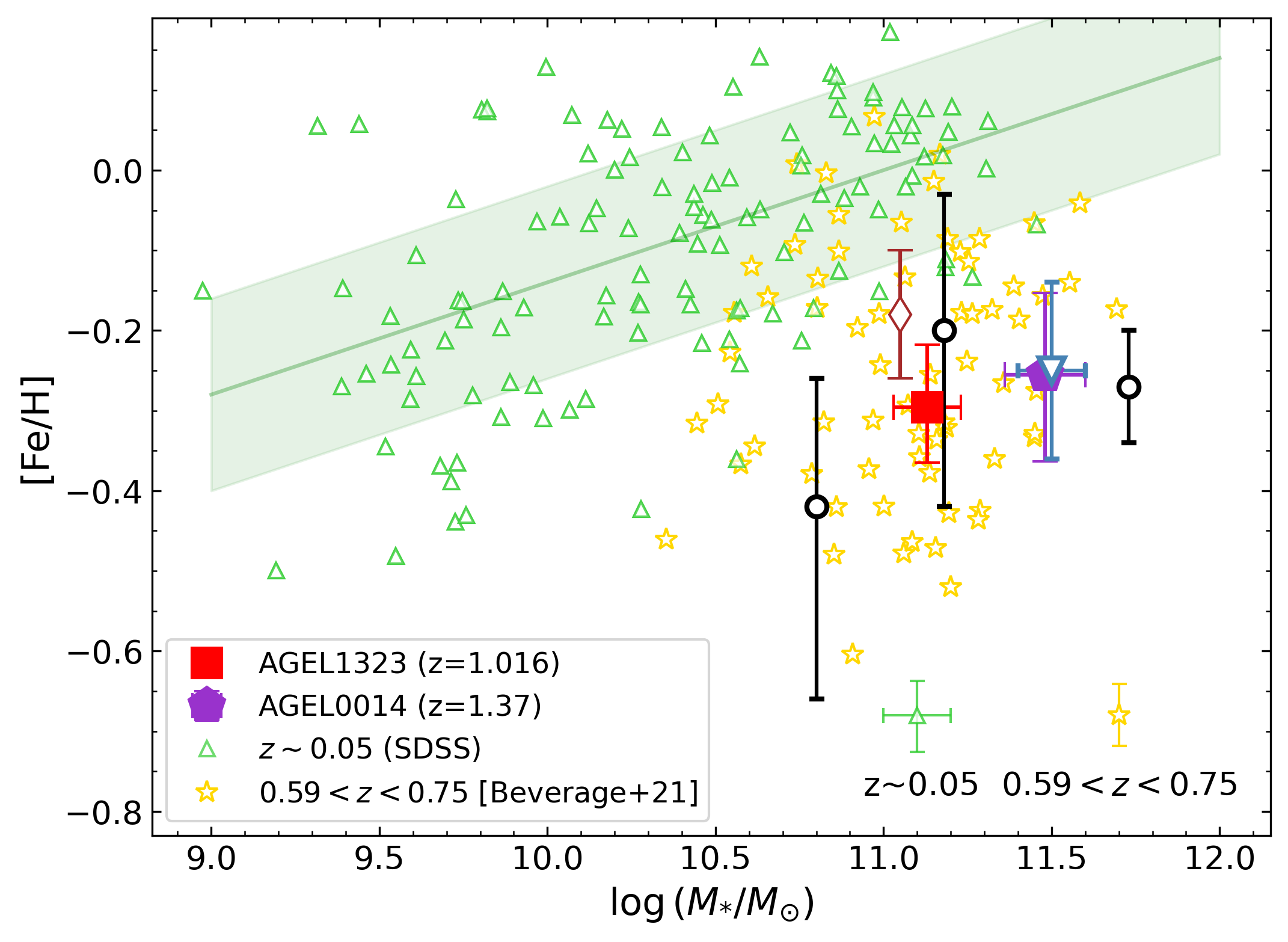}
    \includegraphics[width=0.4\textwidth]{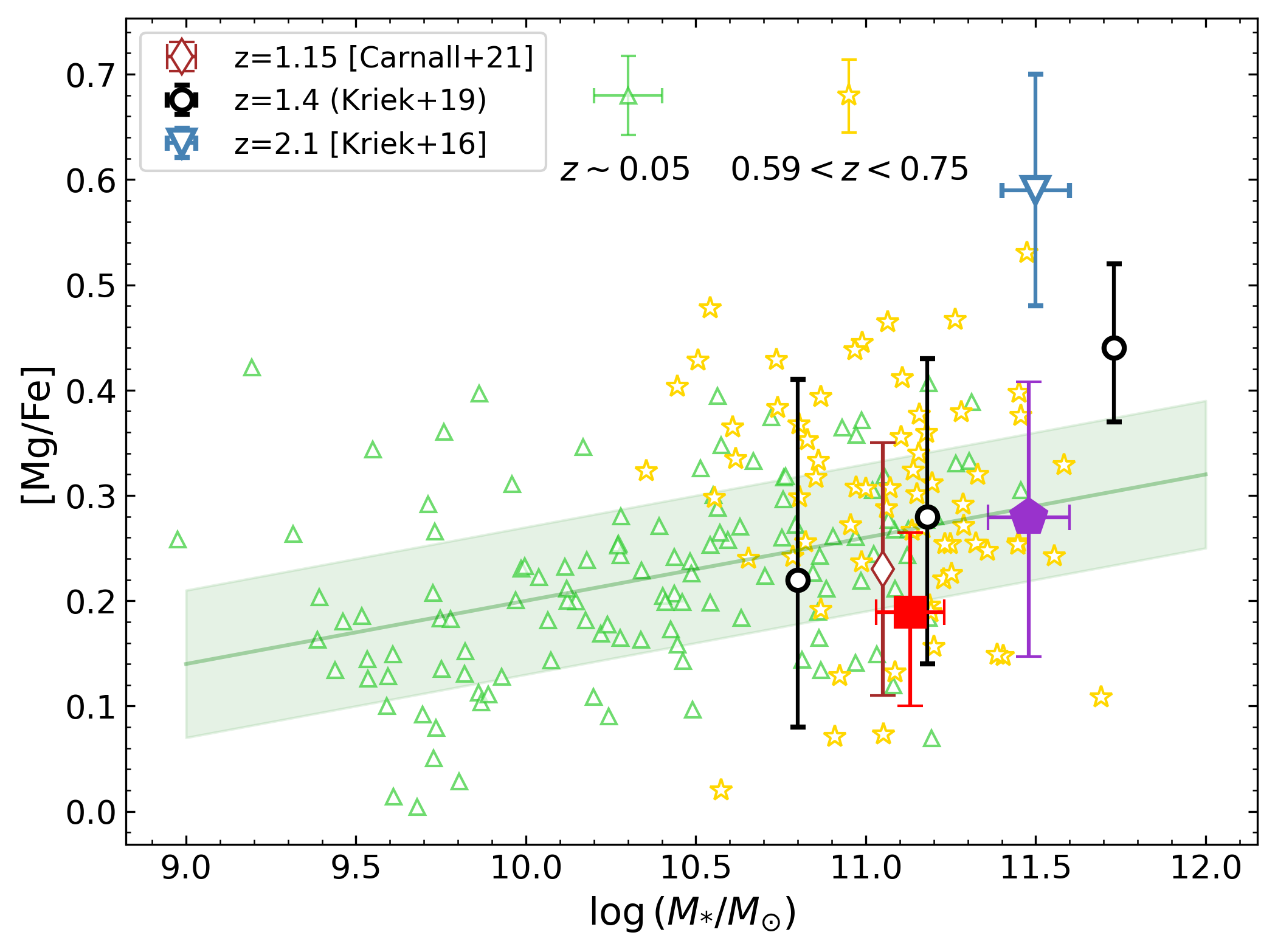}
        \includegraphics[width=0.4\textwidth]{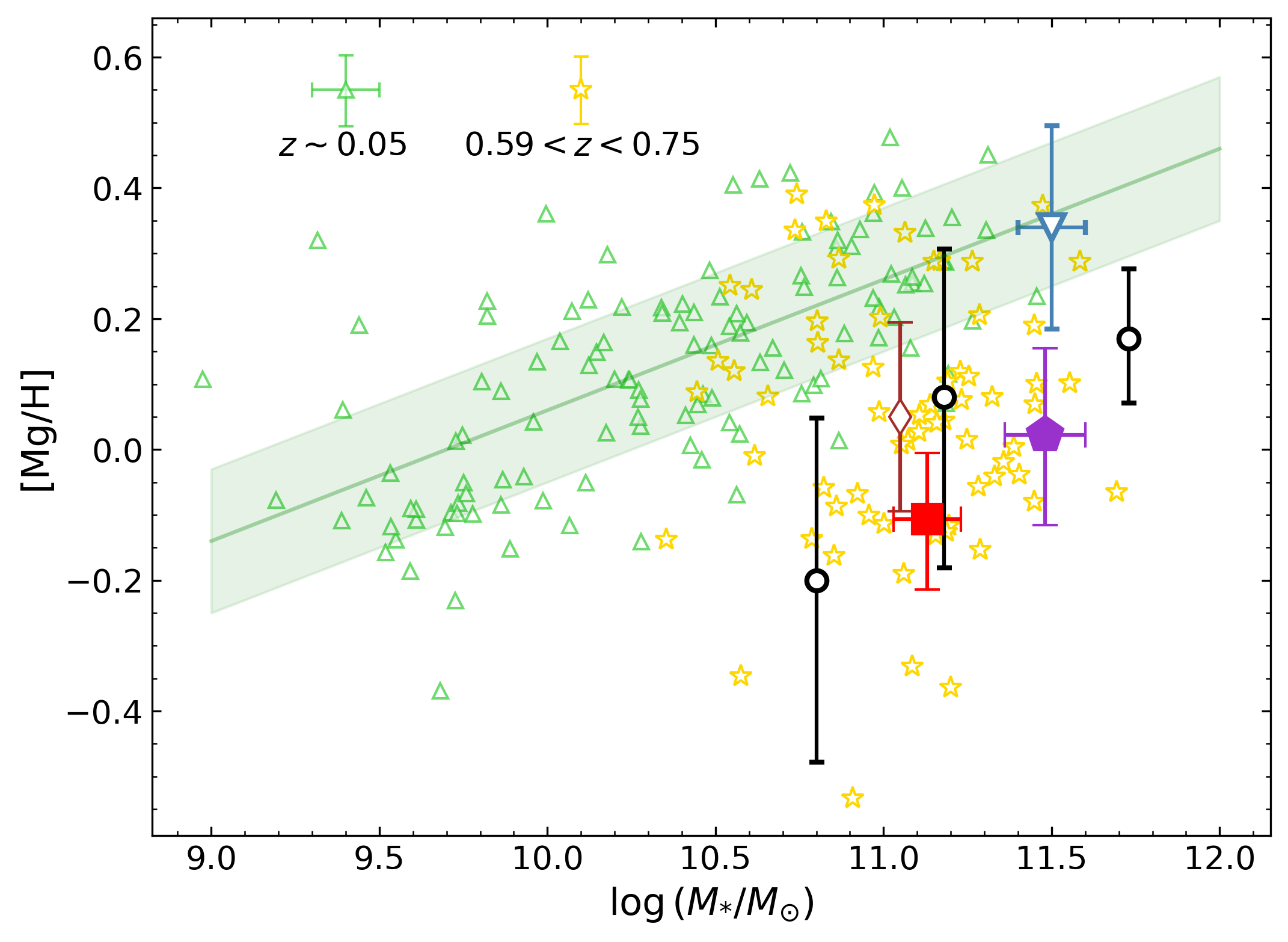}
    \caption{The [Fe/H] (top), [Mg/Fe] (middle) and [Mg/H] (bottom) as a function of stellar mass for quiescent galaxies. \rs\@ at $z=1.016$ and \reb\@ at $z=1.37$  are plotted as a filled square and a pentagon, respectively. The SDSS galaxies at $z\sim 0.05$ (open green triangles) are taken from \citet{Leethochawalit2019} but re-measured with \alf\@ for consistency. The open yellow stars show the measurements for the LEGA-C sample at $0.59<z<0.75$ by \citet{Beverage21}. The abundances of individual high-$z$ quiescent galaxies measured by \citet{Kriek16} (blue open down triangle) and \citet{Kriek19} (black open circles) as well as that of a stacked spectrum at $z=1.15$ determined by \citet{Carnall22} are plotted for comparison. The green solid lines and shaded regions represent the best-fit linear relations and intrinsic scatter of the $z\sim 0.05$ SDSS galaxies.}
        \label{fig:mzr}
\end{figure}

In Figure~\ref{fig:mzr}, we show the measured [Fe/H], [Mg/Fe], [Mg/H] of the \agel\ galaxies compared with other quiescent galaxies at different redshifts as a function of stellar mass. To minimize the systematic effects between various full-spectrum fitting algorithms based on different model assumptions, we limit our comparison to those derived using the \alf\ simple mode. We compare our results with the LEGA-C sample at $z\sim 0.7$ \citep{Beverage21} and other quiescent galaxies at $z\gtrsim 1$ \citep{Kriek16,Kriek19, Carnall22}. We also plot a sample of local SDSS galaxies ($z\sim 0.05$) as reference. The SDSS sample is taken from \citet{Leethochawalit2019}, who randomly selected 152 quiescent galaxies from the \citet{Gallazzi05} sample in the mass range $10^9-10^{11.5}M_{\odot}$. We re-measured their abundances with \alf\ using the same wavelength range as the spectra of the lensed galaxies for consistency. We excluded the galaxies for which posterior distributions hit the upper or lower limits of the priors, leaving a sample of 123 galaxies. The best-fit $z\sim 0$ MZRs of [Fe/H] [Fe/H], [Mg/Fe], and [Mg/H] for the updated \alf\ measurements of nearby SDSS galaxies are: 
\begin{equation}
\begin{aligned}
    {\rm [Fe/H]} & = (-0.14\pm 0.01) + (0.14\pm 0.02)M_{10} \\
    {\rm [Mg/H]} & = (0.06\pm 0.01) + (0.20\pm 0.02)M_{10} \\
    {\rm [Mg/Fe]} & = (0.20\pm 0.01) + (0.06\pm 0.01)M_{10}, \\
\end{aligned}\label{eq:MZR_local}
\end{equation}
where $M_{10} = \log{[M_*/10^{10}M_{\odot}]}$. These relations are shown in Figure~\ref{fig:mzr} to show the evolution in the mass--metallicity relation for the samples at different redshifts.

The lensed galaxies and most of the high-$z$ galaxies are more metal-poor than their local counterparts with similar stellar masses, consistent with the redshift evolution of the MZR found by \citet{Choi14} and \citet{Leethochawalit2019}. However, the [Fe/H] of the \agel\ galaxies, along with other high-$z$ galaxies, appear to be comparable to those of the LEGA-C $z\sim 0.7$ galaxies. This finding does not necessarily contradict the redshift evolution. \citet{Leethochawalit2019} estimated that the normalization of the MZR decreases 0.04~dex per observed Gyr, which translates to a decrease of 0.2~dex in [Fe/H] from $z\sim 0$ to $z\sim 0.7$, and roughly 0.1~dex from $z\sim 0.7$ to $z\sim 1.4$. If we assume the intrinsic scatter of the MZR \cite[$\sim 0.06$~dex at $z\lesssim 0.54$,][]{Leethochawalit2019} has negligible redshift evolution, the combined effects of the observed uncertainties and intrinsic scatter in [Fe/H] at a given stellar mass can account for why the \agel\ galaxies have similar [Fe/H] as LEGA-C galaxies at lower redshifts. A larger sample would be necessary to further investigate the redshift evolution of the MZR at high-$z$. 

\citet{Jafariyazani20} found one quiescent galaxy at $z\sim 2$ to be more metal-rich ($\rm[Fe/H]=0.26$)  than most quiescent galaxies at any redshift. Because this measurement is a significant outlier in metallicity compared to most of the previous studies, which discovered more metal-poor quiescent galaxies at higher redshifts, we do not include the result by \citet{Jafariyazani20} in our comparison.
It is beyond the scope of this work to understand why that galaxy at $z\sim 2$ is significantly more metal-rich than other high-$z$ quiescent galaxies.

\begin{figure}[t]
    \centering
    \includegraphics[width=0.42\textwidth]{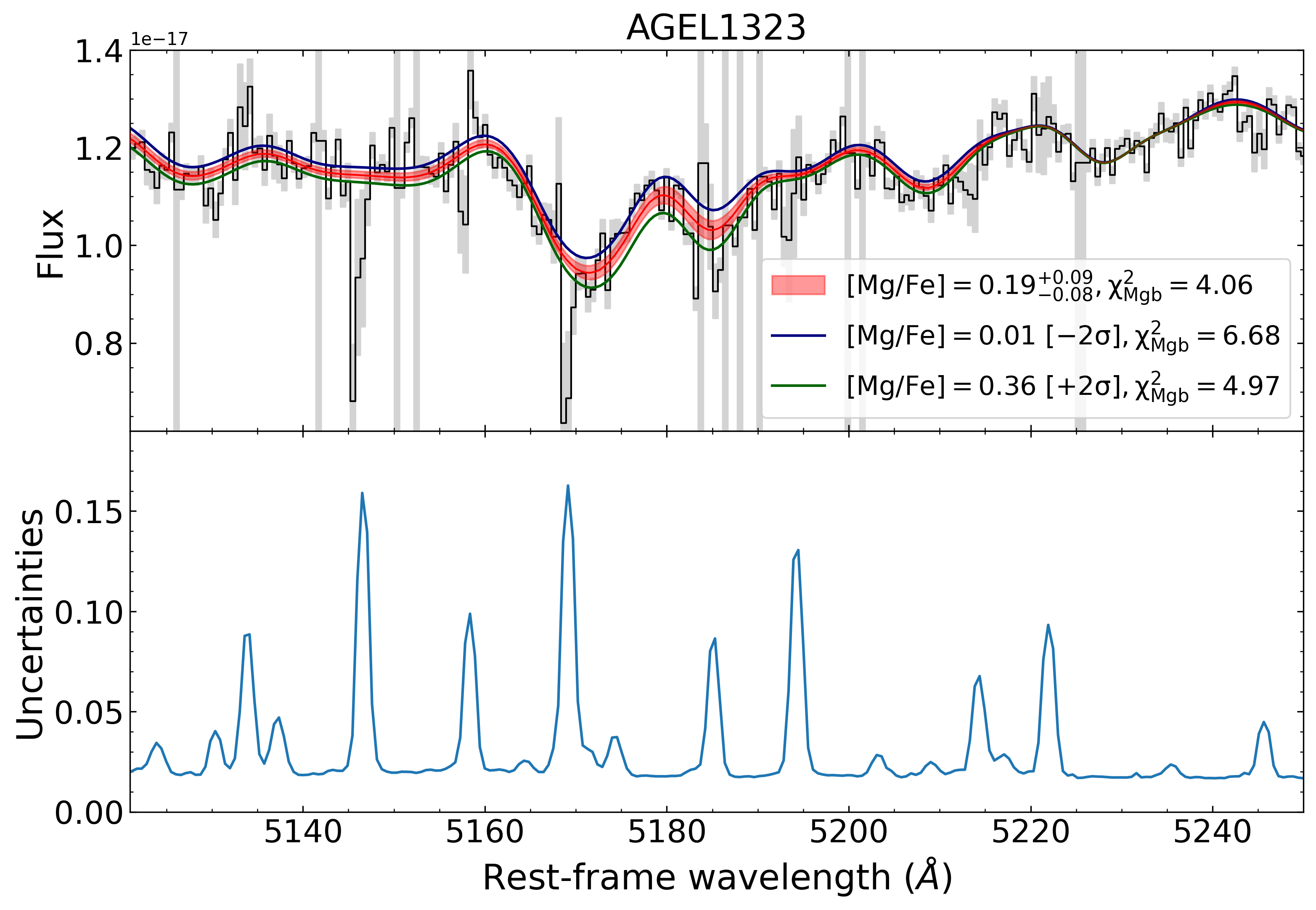}
    \includegraphics[width=0.42\textwidth]{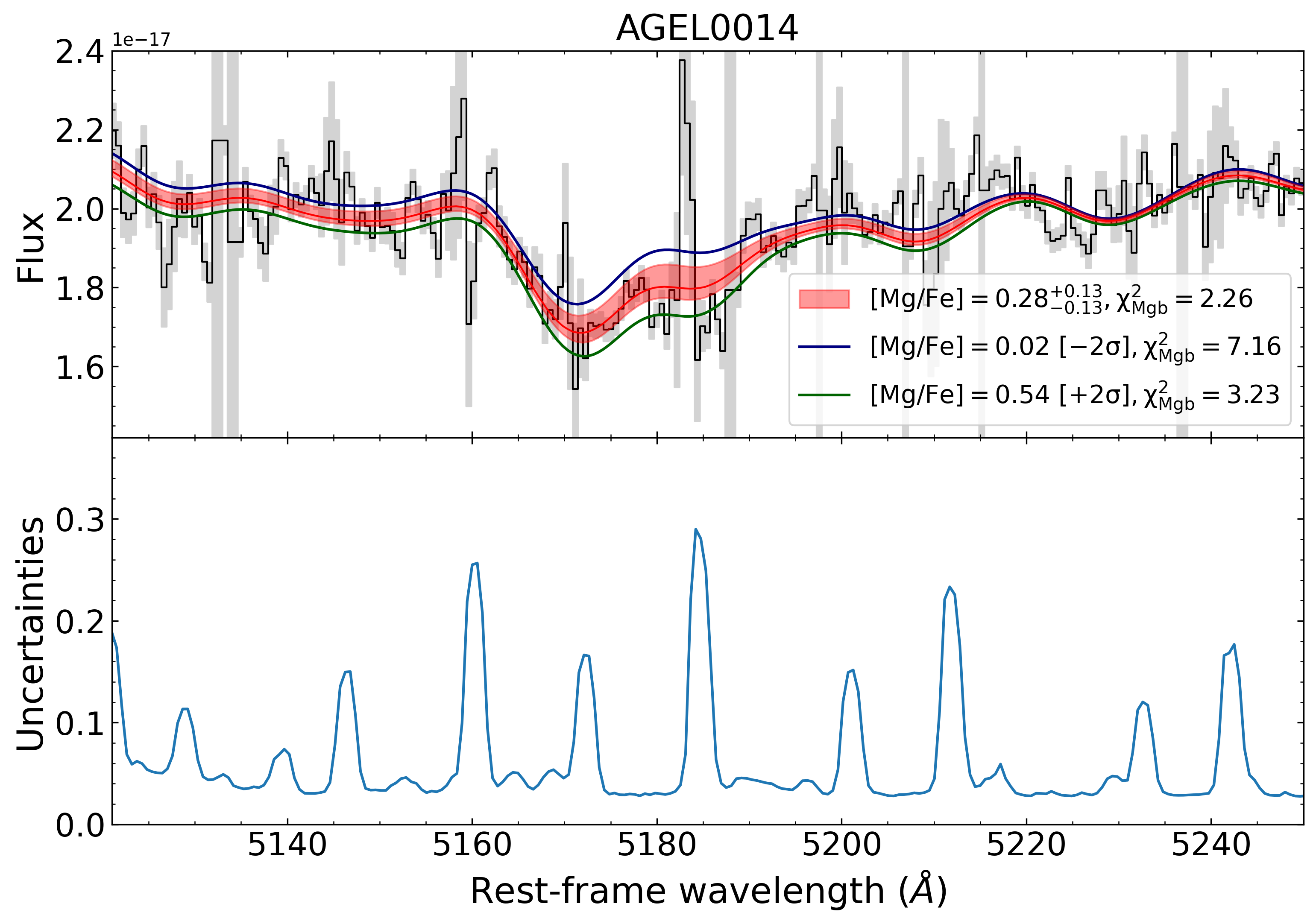}
    \caption{The model spectra with different [Mg/Fe] vs.\ the observed spectra for \rs\ (top) and \reb\ (bottom) near the Mg~b triplet at $\sim$5170~\AA. In each plot, the upper panel shows the models corresponding to 68\% confidence interval of the best-fit [Mg/Fe] by the red region. The model spectra at $2\sigma$ deviation from the median of [Mg/Fe] are overplotted by the blue and green lines for comparison.  The masked pixels are marked by the grey vertical regions. The $1\sigma$ random errors of the observed spectra are indicated by the grey-shaded regions. The lower panel of each plot is the error spectrum, representing the sky background. The models with higher or lower [Mg/Fe] give poorer fits to the Mg~b triplet, which validates our measured [Mg/Fe] from \alf\@.}
        \label{fig:alf_mgb_fit}
\end{figure}

The measurements of [Mg/Fe] of the \agel\ galaxies are comparable to those in the nearby universe at similar masses, although the scatter in the mass--[Mg/Fe] relations of LEGA-C galaxies makes it hard to know if there is any redshift evolution between $z\sim 0$ and $z\sim 0.7$. 
Typically, [Mg/Fe] can be used to probe the SFH\@. Because iron has a more delayed production timescale than magnesium, \citet{Kriek16,Kriek19} used the measured [Mg/Fe] as a proxy for star formation duration when assuming a closed box model, suggesting that high-$z$ quiescent galaxies with enhanced [Mg/Fe] should have shorter star formation timescales. Therefore, the redshift evolution in the mass--[Mg/Fe] relation may reveal whether high-$z$ galaxies have SFHs distinct from those at lower redshifts.

However, the existing studies of redshift evolution in the mass--[Mg/Fe] relation are quite controversial. \citet{Leethochawalit2019} found that [Mg/Fe] is smaller at higher redshifts at fixed stellar mass for cluster quiescent galaxies below $z \lesssim 0.54$, while \citet{Kriek19} and \citet{Beverage21} did not detect any significant redshift evolution for LEGA-C galaxies at $z\sim 0.7$ and three quiescent galaxies at $z\sim 1.4$. Given the current sample size, the mass--[Mg/Fe] relation appears to be similar at local and $z\gtrsim 1$ and thus in favor of what \citet{Beverage21} and \citet{Kriek19} found, but again a larger sample would be needed for stronger constraints. 

The absence of significant $\alpha$-enhancement of the lensed galaxies is unexpected because we expect old, quiescent galaxies to have enhanced [$\alpha$/Fe] ratios. The unexpected result inspired us to consider carefully whether the measured [Mg/Fe] from \alf\ might be underestimated due to imperfect flux calibrations. As a full-spectrum fitting algorithm, \alf\ adopts the information from the whole spectrum, including the continuum and metal absorption lines, to constrain the Mg abundance. \citet{Conroy2018} indicated that the continuum between 4000~\AA\@ and 5300~\AA\@ is also sensitive to Mg abundance, for which any small mismatch between the continuum level of the model spectrum and that of the observed data due to imperfect flux calibrations over a wider wavelength range may bias the measurement of [Mg/Fe]. For each galaxy, we generated \alf\ model spectra with different [Mg/Fe] but fixed all other parameters to the best-fit values returned by \alf\@.

Figure~\ref{fig:alf_mgb_fit} shows  model spectra with [Mg/Fe] varying within $2\sigma$ of the best-fit value compared to the observed data near the Mg~b triplet region for each galaxy. We chose the Mg~b triplet because it is the most prominent Mg feature in the observed wavelength range. These narrow atomic features are less affected by the flux calibrations than wider Mg-bearing molecular bands nearby. We evaluate the reduced $\chi^2$ near the Mg~b for model spectra with $2\sigma$ higher or lower [Mg/Fe] than the best-fit value and compare them with that of the best-fit model. As shown in Figure~\ref{fig:alf_mgb_fit}, for both lensed galaxies, neither a more $\alpha$-rich nor a more $\alpha$-poor model can better describe Mg~b triplet than the best-fit model does, which validates our measured [Mg/Fe] obtained from the full-spectrum fitting.

We now scrutinize the expectation that [Mg/Fe] should be enhanced in old, quiescent galaxies.  In fact, the shape of SFH, merger history, the delayed explosion time of Type Ia supernovae and the presence of outflows/inflows can all affect the final [Mg/Fe] at the time when the galaxies quenched. For this reason, a galaxy that rapidly built its mass (e.g., over $\sim 1$~Gyr) does not necessarily have to be very $\alpha$-enhanced. We have seen such evidence in the Local Group. \citet[][among others]{delosReyes22} found that Sculptor, a dwarf spheroidal galaxy in the Local Group, finished forming stars within $\sim$0.9~Gyr, but has only a moderate $\alpha$-enhancement of ${\rm [Mg/Fe]} \approx 0.2$, comparable to [Mg/Fe] of our lensed galaxies.

In addition, [Mg/Fe] of the lensed galaxies appears to be positively correlated with stellar mass, consistent with the findings of \citet{Kriek19}. We believe future \textit{James Webb Space Telescope} (JWST) observations would be able to yield more reliable measurements of [Mg/Fe] in order to constrain the mass--[Mg/Fe] relation in the high-$z$ universe \citep{Nanayakkara22}.

With the new measurements of our lensed galaxies, a mass dependence of [Mg/H] emerges for quiescent galaxies at $z\gtrsim 1$, suggesting that [Mg/H] is likely to be correlated with stellar mass in the high-$z$ universe. The \agel\ galaxies and other high-$z$ galaxies appear to have lower [Mg/H] compared to local quiescent galaxies. This redshift evolution in the mass--[Mg/H] relation is contradictory to the results of \citet{Leethochawalit2019}, who argued for universal relations for quiescent galaxies at $z\lesssim 0.54$. While the discrepancy may result from different full-spectrum fitting algorithms used to determine [Mg/H], it is more likely that galaxies observed at $z>1$, especially those that formed at $z>3$, may have experienced qualitatively different physical processes shaping the final [Mg/H]. We discuss one of the possible origins in Section~\ref{subsec: mass_loading}.

\section{Discussion}\label{sec:discussion}

\subsection{Dependence of Abundances on Galaxy Formation Time}\label{subsec:dependence_formation_time}

\begin{figure*}
    \centering
\includegraphics[width=0.98\textwidth]{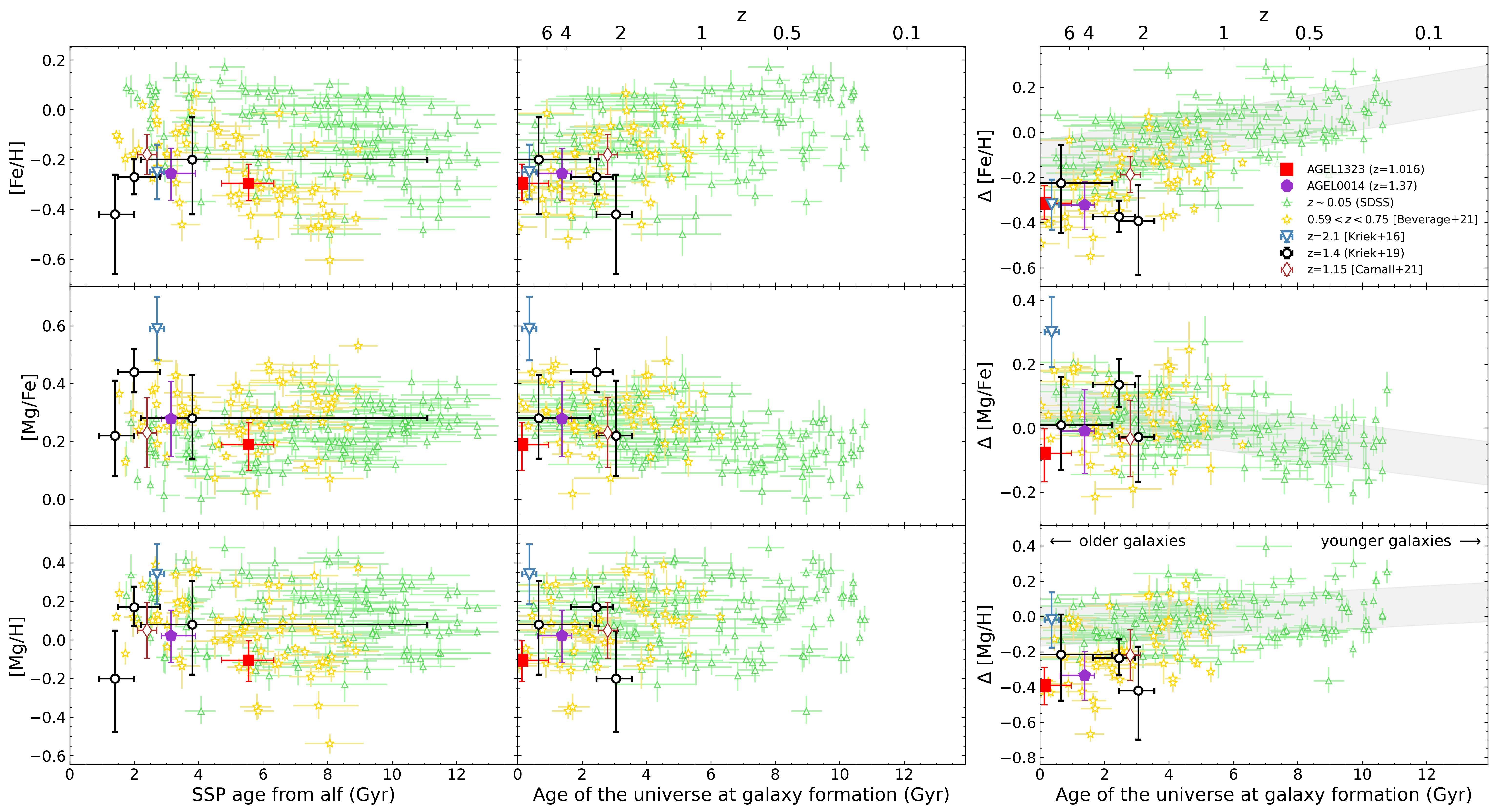}
    \caption{\textit{Left:} Dependence of [Fe/H], [Mg/Fe], and [Mg/H] on the SSP age (first column) and galaxy formation time (second column). \textit{Right}: Dependence of abundances on the galaxy formation time with the stellar mass dependence removed (see Section~\ref{subsec:dependence_formation_time}). The grey shaded regions show the 68\% confidence interval for $z\sim 0$ SDSS galaxies. The symbols for galaxies at different redshifts are the same as in Figure~\ref{fig:mzr}.} 
    \label{fig:formation_time_abundance}
\end{figure*}

As many literature studies have reported the evolution of the MZR with observed redshift for quiescent galaxies \citep[e.g.,][]{Gallazzi14, Choi14, Leethochawalit2019}, it is natural to ask whether such evolution is driven by galaxies' observed redshifts (age of the universe when galaxies were observed) or formation redshifts  (age of the universe when galaxies formed). In other words, should we expect galaxies located at similar redshifts or those forming stars at similar time to have similar metal abundances? \citet{Leethochawalit2019} studied the evolution of the MZR below $z\lesssim 0.54$ and argued that evolution of metallicity is more fundamental with {\it formation} redshift when mass dependence is removed. Following \citet{Leethochawalit2019}, we investigate the dependence of abundances on galaxy formation time here.

We first examine the measured [Fe/H], [Mg/Fe] and [Mg/H] as a function of SSP age obtained from \alf's simple mode for quiescent galaxies at various redshifts. As shown in Figure~\ref{fig:formation_time_abundance}, [Fe/H] of high-$z$ quiescent galaxies are distinct from those of local SDSS quiescent galaxies at a given SSP age, while such significant difference is not seen in [Mg/H]. At a certain SSP age, high-$z$ quiescent galaxies are relatively more metal-poor and more $\alpha$-enhanced compared to local counterparts. This scenario is expected as high-$z$ galaxies have less time for Type Ia supernova explosion for iron production. The lack of prominent [Mg/H] separation between local and high-$z$ populations can be explained by the short recycling timescale of magnesium. The distributions of [Mg/Fe] for high-z and local populations are slightly different, as the production of iron and magnesium both affect this trend.

If high-$z$ galaxies evolve passively and remain quenched to $z\sim 0$, their stellar abundances would not change significantly due to the absence of new star formation. Because high-$z$ quiescent galaxies are believed to be the progenitors of local quiescent galaxies, we expect the high-$z$ populations to overlap with the local counterparts in age--abundance plots once the age differences are corrected. Here, we approach this problem in a reverse way. If we ignore the small metallicity change as galaxies evolve passively, we can correct the age differences between the high-$z$ and local galaxies by tracing back the age of the universe at galaxy formation, which is the difference between the age of the universe at the observed redshift and the measured SSP age for each galaxy. As can be seen in the second column of Figure~\ref{fig:formation_time_abundance}, the differences of [Fe/H] and [Mg/Fe] between high-$z$ galaxies and local galaxies are attenuated when we compare them as a function of galaxy formation time rather than SSP age.

We further remove the first-order mass dependence of abundances by subtracting the best-fit MZRs at $z\sim 0$ (Equation~\ref{eq:MZR_local}) from the observed abundances. The deviations in the right panels of Figure~\ref{fig:formation_time_abundance} demonstrate how galaxy formation time affects the final abundances. At $z\sim 0$, the evolution in [Fe/H] and [Mg/Fe] with formation time is $0.024\pm 0.003$~dex per Gyr and $-0.013\pm 0.002$~dex per Gyr, respectively. On the contrary, we do not detect a significant dependence of [Mg/H] on galaxy formation time for local galaxies, with the best-fit slope of $0.010\pm 0.004$~dex per formation Gyr. The weak evolution of [Mg/H] with formation age is in agreement with the short recycling timescale of magnesium. As can be seen in Figure~\ref{fig:formation_time_abundance}, the LEGA-C and high-$z$ galaxies exhibit similar evolution of abundances when the dependence on mass and observed redshift is removed. The tighter relations in [Fe/H] and [Mg/Fe] imply that at a fixed stellar mass, the galaxy formation time since the Big Bang plays more fundamentally determines stellar metallicity and $\alpha$-enhancement than the observed redshift.

Still, [Fe/H] and [Mg/H] of the LEGA-C galaxies and high-$z$ galaxies are systematically lower than local galaxies when the dependence on mass and age is removed, although the differences are relatively small. The discrepancy for high-$z$ galaxies implies that additional mechanisms are needed to explain the small offset. 

One possibility is that not all high-$z$ quiescent galaxies evolve passively. There is a chance that some of them may rejuvenate and form younger stars in a later epoch. Recent studies have suggested that accretion of satellite galaxies is important to explain the size growth of massive elliptical galaxies across cosmic time  \citep[e.g.,][and references therein]{Oser2010,Greene13,Oyarzun19}. Such mergers may trigger star formation in originally quenched galaxies. 
If some of these high-$z$ quiescent galaxies were at the temporary quenching phase between two starbursts, they would have a higher [Fe/H] and a younger formation time when reaching $z\sim 0$. We believe this scenario may be necessary to explain the redshift evolution of the MZR in Figure~\ref{fig:mzr}. As high-$z$ quiescent galaxies evolve to $z\sim 0$ passively, they would become less massive and slightly more metal-poor. It appears that the most massive galaxies cannot move to the local sequence unless they can get metal-enriched in the later epoch, which would require either new star formation or mergers with much more metal-rich systems. The latter is less likely to happen because massive galaxies tend to accrete more metal-poor satellite galaxies \citep{Oser2010}.

With the few available measurements at $z\gtrsim 1$, we conclude that galaxy formation time would still be the key factor determining the metal abundances in quiescent galaxies below $z\lesssim 2$.  At the same time, it is possible that some of the high-$z$ galaxies may re-start forming stars at a later epoch. A larger sample of quiescent galaxies at $z\gtrsim 1$ would be necessary to confirm this possible rejuvenation in star formation.

\subsection{Enhanced outflows during the star formation}\label{subsec: mass_loading}

\begin{figure}[t]
    \centering
    \includegraphics[height=2.4in]{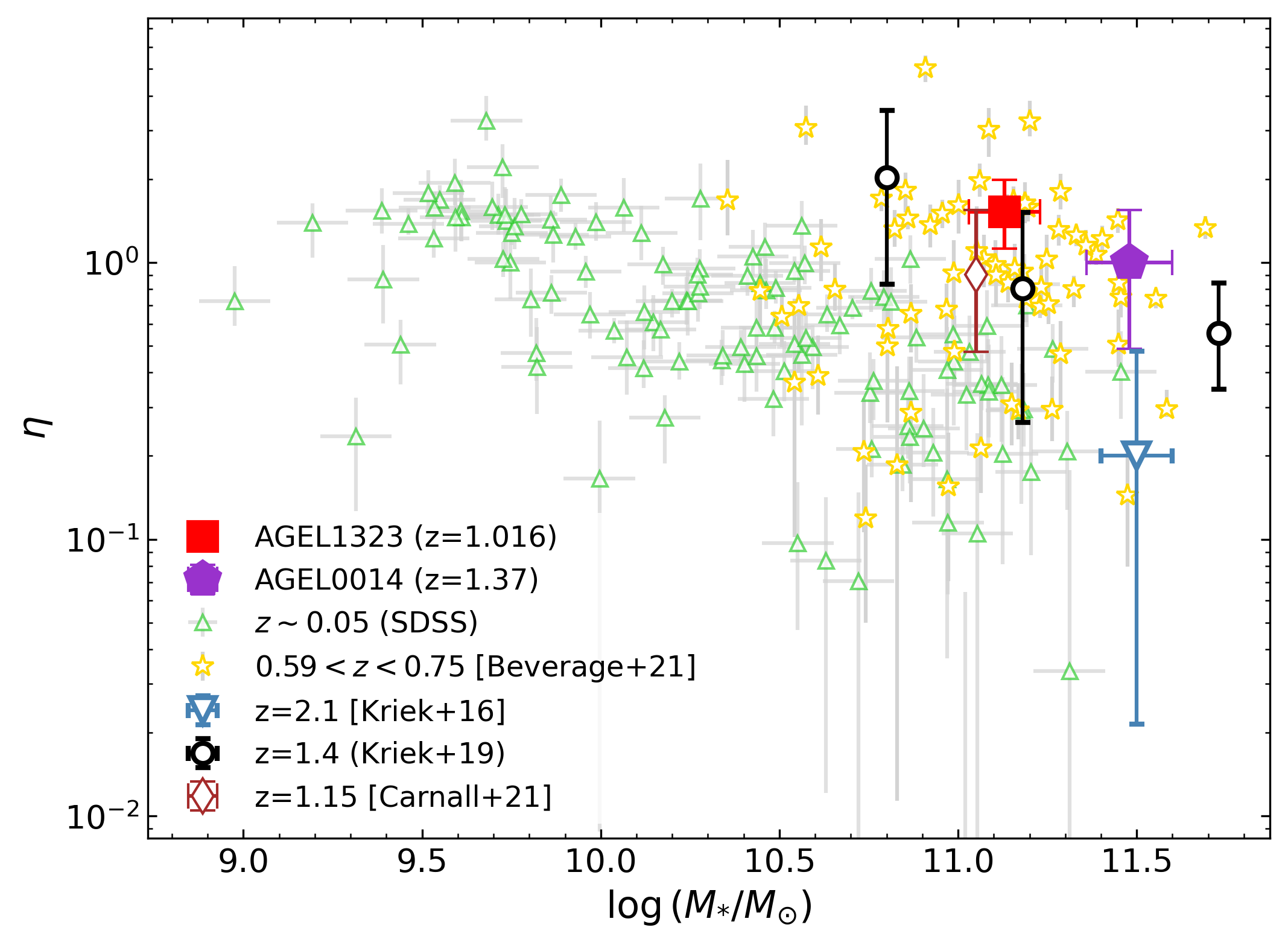}
    \caption{The mass-loading factor ($\eta$) inferred from the measured [Mg/H] as a function of stellar mass for quiescent galaxies. The sample and the symbols are the same as Figure~\ref{fig:mzr}. The two \agel\@ galaxies appear to have higher mass-loading factors than nearby galaxies at similar masses.  } 
    \label{fig:mzr_eta}
\end{figure}

To understand why \rs\ and \reb\ quenched so early in the universe, we use the measured [Mg/H] to constrain the outflows of the lensed galaxies during their star formation. \citet{Leethochawalit2019} presented a simple analytic chemical evolution model for quiescent galaxies that connects instantaneously recycled metals to the time-averaged mass-loading factor, based on the work of \citet{Lu15}. Readers are encouraged to read \citet{Leethochawalit2019} for the details of the model. To summarize, the mass-loading factor here is defined as the ratio of the mass outflow rate to the SFR, averaged over the entire SFH\@. They also assume that: 
\begin{enumerate}
    \item The interstellar medium (ISM) where star formation occurs is perfectly mixed. 
    \item Metals are instantaneously recycled. 
    \item Outflows and inflows are permitted, but only the outflows can significantly affect the total metal budget in the ISM.  
\end{enumerate}
Because magnesium is the product of core-collapse supernovae of short-lived stars, it can be approximated as an instantaneously recycled element and thus fit into the model above. If assuming that no or minimal amount of gas is left in the quiescent galaxies, the abundance of an instantaneously recycled element would be 
\begin{equation}
Z_{*,\rm quiescent} \approx \frac{y}{1+\frac{\langle \eta \rangle}{1-R}},
\label{eq:mass_loading}
\end{equation}
where $y$ is the supernova yield and $R$ is the return mass fraction defined to be the fraction of the mass of a stellar generation that returns to the interstellar medium from short-lived massive stars and stellar winds. $Z_{*,\rm quiescent}$ can be substituted as the absolute Mg abundance of quiescent galaxies. Following \citet{Leethochawalit2019}, we adopted the solar abundance of Mg from \citet{Asplund09} and the yield as three times the solar Mg abundance \citep{Nomoto06} to calculate the absolute Mg abundance.  The return fraction is set to $R=0.46$ \citep{Lu15} for all galaxies.

In Figure~\ref{fig:mzr_eta}, we show the mass-loading factors inferred from the measured [Mg/H] for quiescent galaxies at various redshifts. The \agel\ galaxies---as well as other high-$z$ galaxies---appear to have higher mass-loading factors than nearby quiescent galaxies, implying that galaxies may have enhanced outflows when they were forming stars in the early universe. The strong outflows would naturally cause the two galaxies to lose gas quickly and thus can no longer sustain further star formation. 

The observational constraints on the mass-loading factors for high-$z$ galaxies are qualitatively consistent with the FIRE simulations \citep{Hayward17}, which predict that $\eta$ increases significantly as redshift increases at $M_* \gtrsim 10^{10}M_{\odot}$. 
Therefore, it will be essential to perform similar studies when a larger sample of high-$z$ quiescent galaxies at $z>1$ is available in order to further investigate whether there is a redshift evolution in the mass-loading factor as suggested by the FIRE simulations.

\subsection{Comparison with \citet{Sukay22}}
\citet{Sukay22} derived the stellar population properties of \rs\ via a joint fit of photometry and spectroscopy with \texttt{Prospector} \citep{Johnson21}. They adopted a non-parametric SFH model with seven age bins to characterize the stellar population. In this work, we instead measure the stellar population via full-spectrum fitting algorithms \alf\ assuming an SSP or two SSPs (i.e., one or two age bins). \citet{Sukay22} used the spectrum obtained on Alhambra Faint Object Spectrograph and Camera (ALFOSC) at the 2.56~m Nordic Optical Telescope (NOT) with a total integration time of 80~mintues. The low-resolution spectrum ($R=590$) covers the rest-frame wavelength between 2820~\AA\ and 5030~\AA\ and thus is unable to resolve most of the faint metal absorption lines which is important to robustly constrain the stellar metallicity. 

Our new Keck spectra therefore are much deeper and have a much higher resolution. The Keck spectra cover the rest-frame optical between 3600-5500 \AA, which includes many more metal absorption lines essential to the determination of [Fe/H] and [Mg/Fe]. We prefer an SSP model over more complicated SFHs because its simplicity reduces the degeneracy among free parameters. Although we adopted a different SFH model, the SSP age in this work is still in qualitatively good agreement with the SFH derived by \citet{Sukay22}. We obtained an SSP age of $5.6\pm 0.8$~Gyr, while the SFH derived by \citet{Sukay22} indicates an absence of prominent star formation activity until $\sim$3~Gyr ago. Both results indicate that \rs\ is an old system that quenched very early in the universe. 

We measured elemental abundances of $\rm [Fe/H]=-0.30^{+0.08}_{-0.07}$ and $\rm [Mg/Fe]=0.19^{+0.08}_{-0.09}$. If we convert our measurements to total metallicity as ${\rm [Z/H]} = {\rm [Fe/H]} + 0.94{\rm [Mg/Fe]}$ \citep{Thomas03}, the total metallicity is $\rm [Z/H]=-0.12\pm 0.11$, which is consistent with \citeauthor{Sukay22}'s best-fit stellar metallicity, $\log(Z/Z_{\odot})=-0.19$.

\subsection{Comparison between the ages and metallicities derived from different approaches}

While we primarily focus on the ages and the metallicities derived from \alf\ with full-spectrum fitting, we also measured the stellar populations with SED fitting to estimate the stellar masses (Section~\ref{sec:SED_fitting_mass}). Here we briefly discuss the differences in the measurements from the two approaches.

Since we adopt the delayed-$\tau$ model in the broadband SED fitting, we compare the mass-weighted stellar ages averaged over the SFHs with the SSP ages derived from \alf\@. As shown in Figure~\ref{sec:SED_fitting_mass}, we obtained a mean stellar age of $3.6^{+0.8}_{-1.1}$~Gyr for \rs\ and $2.3^{+0.7}_{-0.9}$~Gyr for \reb\@. Although the ages derived from broadband SED fitting are still consistent with the SSP ages within $2\sigma$, they are systematically lower than the values measured with \alf\@. The mean stellar ages vary slightly across different SFH models, but all of them predict ages younger than the SSP ages while remaining consistent with \alf\@'s measurements within $2\sigma$. Even if such age differences can be attenuated by conducting a joint fitting of photometry and spectroscopy with BAGPIPES, the measured ages of both galaxies are still lower than those measured from \alf\@.

As for the stellar metallicities, we determine the [Z/H]$_{\rm BAGPIPES}= -0.20^{+0.07}_{-0.08}$ for \rs\ and [Z/H]$_{\rm BAGPIPES}=-0.17^{+0.03}_{-0.04}$ for \reb\ when the red spectrum covering the Mg~b triplet and Fe~I lines is included for each galaxy. The delayed-$\tau$ SFH with a single metallicity is assumed. The stellar metallicities measured from the joint fitting of photometry and spectroscopy are therefore consistent with the [Fe/H] determined by \alf\ within $1\sigma$, although BAGPIPES still yields higher metallicity measurements. The BAGPIPES measurement of [Z/H] of \rs\ is also within $1\sigma$ of the total metallicity when we convert the \alf's measurements of [Fe/H] and [Mg/Fe] into [Z/H] as ${\rm [Z/H]} = {\rm [Fe/H]} + 0.94{\rm [Mg/Fe]}$ \citep{Thomas03}, for consistency with other recent work \citep[e.g.,][]{Kriek19, Carnall19}. However, the total metallicities of \reb\ are consistent between the two approaches within $2\sigma$.

Even though the stellar ages and metallicities measured from SED fitting with BAGPIPES and full-spectrum fitting with \alf\ appear to be consistent with each other, SED fitting systematically recovers younger and more metal-rich populations than full-spectrum fitting does. Several factors may account for the potential systematic effects, such as different SSP templates and the choices of SFH models. The fitted wavelength region may also affect the results. BAGPIPES only allows one spectrum to be included in the fit while \alf\ can analyze two spectra obtained on different instruments/gratings. SED fitting also spans a much wider spectral range than full-spectrum fitting. There is no doubt that different spectral regions are sensitive to stellar populations at different ages. With the current sample of only two galaxies, it is beyond the scope of this work to quantify the possible systematic differences in the stellar population parameters measured with different approaches.

\section{Conclusions}\label{sec:summary}
We presented the deep rest-frame optical spectra obtained on the Keck~I telescope for two gravitationally lensed, massive quiescent galaxies at $z\gtrsim 1$: \rs\ ($\log{(M_*/M_{\odot})}
 \sim 11.1$) and \reb\ ($\log{(M_*/M_{\odot})} \sim 11.5$). The high S/N of the spectra enabled us to robustly characterize the stellar population and elemental abundances of the lensed galaxies via full-spectrum fitting under the SSP assumption. A summary of our findings is as follows:

\begin{enumerate}

    \item 
    We determined the stellar metallicities as ${\rm [Fe/H]}=-0.30^{+0.08}_{-0.07}$ for \rs\ ($z=1.016$) and ${\rm [Fe/H]}=-0.26^{+0.10}_{-0.11}$ for \reb\ ($z=1.374)$. Compared with low-$z$ galaxies at similar masses, the lensed galaxies in our sample are more metal-poor, consistent with other stellar metallicity measurements of quiescent galaxies at similar redshift \citet{Kriek19, Carnall19}. This tentatively indicates that the redshift evolution of the stellar MZR should be in place at $z\lesssim 1.4$. When we remove the mass dependence of metallicity for quiescent galaxies at different redshifts, we find that galaxies of the same masses that formed at the same time have similar metallicities, regardless of their observed redshift.  This result implies that the evolution of the stellar MZR is more fundamental with formation redshift than with observed redshift. 

    \item
    We also measured the $\alpha$-enhancements for two lensed galaxies as ${\rm [Mg/Fe]}=0.19^{+0.08}_{-0.09}$ for \rs\ and ${\rm [Mg/Fe]}=0.28\pm 0.13$ for \reb\@. The lensed galaxies have comparable [Mg/Fe] as their low-$z$ counterparts, despite of their old ages. 

    \item
    We obtained SSP-equivalent ages of $5.6\pm 0.8$~Gyr for \rs\ and $3.1^{+0.8}_{-0.3}$~Gyr for \reb\@. Considering the observed redshifts and stellar masses of the lensed galaxies, the old ages show that the majority of the stellar population formed very rapidly in the galaxies (less than 2~Gyr after the Big Bang). Using a simple chemical evolution \citep{Leethochawalit2019}, we inferred mass-loading factors of the galaxies from the measured [Mg/H]\@. The high mass-loading factors imply that the two galaxies both experienced enhanced outflows during their star formation, which may have led to their early quenching.
\end{enumerate}

This work has demonstrated how gravitational lensing can further our understanding of the chemical evolution of quiescent galaxies in the high-redshift universe. At the moment, there are still just a handful of robust stellar abundance measurements for quiescent galaxies at $z \gtrsim 1$.  Therefore, larger samples would be necessary to systematically investigate the relation between stellar mass, stellar abundances, and the inferred mass-loading factors of high-$z$ quiescent galaxies. The early data from JWST have revealed numerous new lens candidates that may be quiescent galaxies from even higher redshifts than those presented in this work. The upcoming bounty of spectroscopic and imaging data will revolutionize our knowledge of chemical enrichment histories of quiescent galaxies in the early universe.  

\begin{acknowledgments}

The authors acknowledge the insightful and constructive feedback from the anonymous referee, which helped us improve the manuscript. We would also like to thank Meng Gu for helpful advice on full-spectrum fitting with \alf\@ and Allison Strom for useful discussions on MOSFIRE data reduction and analysis with MOSPEC\@. We gratefully thank the staff at the W. M. Keck Observatory, including support astronomers Luca Rizzi, Carlos Alvarez and Jim Lyke, and telescope operators Arina Rostopchina, Julie Renauld-Kim, and Heather Hershey, for assisting in the observations.

We are grateful to the many people who have worked to make the Keck Telescope and its instruments a reality and to operate and maintain the Keck Observatory.  The authors wish to extend special thanks to those of Hawaiian ancestry, on whose sacred mountain we are privileged to be guests.  Without their generous hospitality, none of the observations presented herein would have been possible. 

This material is based on work supported by the National Science Foundation under Grant No.\ AST-2233781 (E.N.K.\@) and AST-2009278 (C.C.S.\@). Z.Z.\@, E.N.K.\ and C.C.S.\ acknowledge financial support from National Aeronautics and Space Administration (NASA) through the FINESST program (No.\ 80NSSC22K1755). K.G.\@, T.N.\ and C.J.\ acknowledge support from Australian Research Council (FL180100060). S.M.S.\ acknowledges funding from the Australian Research Council (DE220100003). T.J.\ and K.V.G.C.\ gratefully acknowledge financial support from NASA through grant HST-GO-16773, the Gordon and Betty Moore Foundation through Grant GBMF8549, and the National Science Foundation through grant AST-2108515. Parts of this research were conducted by the Australian Research Council Centre of Excellence for All Sky Astrophysics in 3 Dimensions (ASTRO 3D), through project number CE170100013.

This work makes use of observations with the NASA/ESA Hubble Space Telescope obtained from the Mikulski Archive for Space Telescopes at the Space Telescope Science Institute (STScI), which is operated by the Association of Universities for Research in Astronomy, Incorporated, under NASA contract NAS5-26555. Support for Program number GO-16773 was provided through a grant from the STScI under NASA contract NAS5-26555.

\facilities{Keck I (LRIS, MOSFIRE), HST (WFC3) }
\software{\texttt{SExtractor} \citep{Bertin96}, \texttt{alf} \citep{Conroy2018}, \texttt{dynesty} \citep{dynesty}, 
\texttt{PyAutoLens} \citep{Nightingale2015, Nightingale2018, pyautofit, pyautolens},  BAGPIPES \citep{Carnall18, Carnall19}, Astropy \citep{Astropy2013,Astropy2018}, NumPy \citep{2020NumPy-Array}, SciPy \citep{2020SciPy-NMeth}}

\end{acknowledgments}


\bibliography{main}

\begin{thebibliography}{}
\expandafter\ifx\csname natexlab\endcsname\relax\def\natexlab#1{#1}\fi
\providecommand{\url}[1]{\href{#1}{#1}}
\providecommand{\dodoi}[1]{doi:~\href{http://doi.org/#1}{\nolinkurl{#1}}}
\providecommand{\doeprint}[1]{\href{http://ascl.net/#1}{\nolinkurl{http://ascl.net/#1}}}
\providecommand{\doarXiv}[1]{\href{https://arxiv.org/abs/#1}{\nolinkurl{https://arxiv.org/abs/#1}}}

\bibitem[{{Abbott} {et~al.}(2018){Abbott}, {Abdalla}, {Allam}, {Amara},
  {Annis}, {Asorey}, {Avila}, {Ballester}, {Banerji}, {Barkhouse}, {Baruah},
  {Baumer}, {Bechtol}, {Becker}, {Benoit-L{\'e}vy}, {Bernstein}, {Bertin},
  {Blazek}, {Bocquet}, {Brooks}, {Brout}, {Buckley-Geer}, {Burke}, {Busti},
  {Campisano}, {Cardiel-Sas}, {Carnero Rosell}, {Carrasco Kind}, {Carretero},
  {Castander}, {Cawthon}, {Chang}, {Chen}, {Conselice}, {Costa}, {Crocce},
  {Cunha}, {D'Andrea}, {da Costa}, {Das}, {Daues}, {Davis}, {Davis}, {De
  Vicente}, {DePoy}, {DeRose}, {Desai}, {Diehl}, {Dietrich}, {Dodelson},
  {Doel}, {Drlica-Wagner}, {Eifler}, {Elliott}, {Evrard}, {Farahi}, {Fausti
  Neto}, {Fernandez}, {Finley}, {Flaugher}, {Foley}, {Fosalba}, {Friedel},
  {Frieman}, {Garc{\'\i}a-Bellido}, {Gaztanaga}, {Gerdes}, {Giannantonio},
  {Gill}, {Glazebrook}, {Goldstein}, {Gower}, {Gruen}, {Gruendl}, {Gschwend},
  {Gupta}, {Gutierrez}, {Hamilton}, {Hartley}, {Hinton}, {Hislop}, {Hollowood},
  {Honscheid}, {Hoyle}, {Huterer}, {Jain}, {James}, {Jeltema}, {Johnson},
  {Johnson}, {Kacprzak}, {Kent}, {Khullar}, {Klein}, {Kovacs}, {Koziol},
  {Krause}, {Kremin}, {Kron}, {Kuehn}, {Kuhlmann}, {Kuropatkin}, {Lahav},
  {Lasker}, {Li}, {Li}, {Liddle}, {Lima}, {Lin}, {L{\'o}pez-Reyes}, {MacCrann},
  {Maia}, {Maloney}, {Manera}, {March}, {Marriner}, {Marshall}, {Martini},
  {McClintock}, {McKay}, {McMahon}, {Melchior}, {Menanteau}, {Miller},
  {Miquel}, {Mohr}, {Morganson}, {Mould}, {Neilsen}, {Nichol}, {Nogueira},
  {Nord}, {Nugent}, {Nunes}, {Ogando}, {Old}, {Pace}, {Palmese},
  {Paz-Chinch{\'o}n}, {Peiris}, {Percival}, {Petravick}, {Plazas}, {Poh},
  {Pond}, {Porredon}, {Pujol}, {Refregier}, {Reil}, {Ricker}, {Rollins},
  {Romer}, {Roodman}, {Rooney}, {Ross}, {Rykoff}, {Sako}, {Sanchez}, {Sanchez},
  {Santiago}, {Saro}, {Scarpine}, {Scolnic}, {Serrano}, {Sevilla-Noarbe},
  {Sheldon}, {Shipp}, {Silveira}, {Smith}, {Smith}, {Smith}, {Soares-Santos},
  {Sobreira}, {Song}, {Stebbins}, {Suchyta}, {Sullivan}, {Swanson}, {Tarle},
  {Thaler}, {Thomas}, {Thomas}, {Troxel}, {Tucker}, {Vikram}, {Vivas},
  {Walker}, {Wechsler}, {Weller}, {Wester}, {Wolf}, {Wu}, {Yanny}, {Zenteno},
  {Zhang}, {Zuntz}, {DES Collaboration}, {Juneau}, {Fitzpatrick}, {Nikutta},
  {Nidever}, {Olsen}, {Scott}, \& {NOAO Data Lab}}]{Abbott18}
{Abbott}, T.~M.~C., {Abdalla}, F.~B., {Allam}, S., {et~al.} 2018, \apjs, 239,
  18, \dodoi{10.3847/1538-4365/aae9f0}

\bibitem[{{Aihara} {et~al.}(2022){Aihara}, {AlSayyad}, {Ando}, {Armstrong},
  {Bosch}, {Egami}, {Furusawa}, {Furusawa}, {Harasawa}, {Harikane}, {Hsieh},
  {Ikeda}, {Ito}, {Iwata}, {Kodama}, {Koike}, {Kokubo}, {Komiyama}, {Li},
  {Liang}, {Lin}, {Lupton}, {Lust}, {MacArthur}, {Mawatari}, {Mineo},
  {Miyatake}, {Miyazaki}, {More}, {Morishima}, {Murayama}, {Nakajima},
  {Nakata}, {Nishizawa}, {Oguri}, {Okabe}, {Okura}, {Ono}, {Osato}, {Ouchi},
  {Pan}, {Plazas Malag{\'o}n}, {Price}, {Reed}, {Rykoff}, {Shibuya},
  {Simunovic}, {Strauss}, {Sugimori}, {Suto}, {Suzuki}, {Takada}, {Takagi},
  {Takata}, {Takita}, {Tanaka}, {Tang}, {Taranu}, {Terai}, {Toba}, {Turner},
  {Uchiyama}, {Vijarnwannaluk}, {Waters}, {Yamada}, {Yamamoto}, \&
  {Yamashita}}]{Aihara22}
{Aihara}, H., {AlSayyad}, Y., {Ando}, M., {et~al.} 2022, \pasj, 74, 247,
  \dodoi{10.1093/pasj/psab122}

\bibitem[{{Asplund} {et~al.}(2009){Asplund}, {Grevesse}, {Sauval}, \&
  {Scott}}]{Asplund09}
{Asplund}, M., {Grevesse}, N., {Sauval}, A.~J., \& {Scott}, P. 2009, \araa, 47,
  481, \dodoi{10.1146/annurev.astro.46.060407.145222}

\bibitem[{{Astropy Collaboration} {et~al.}(2013){Astropy Collaboration},
  {Robitaille}, {Tollerud}, {Greenfield}, {Droettboom}, {Bray}, {Aldcroft},
  {Davis}, {Ginsburg}, {Price-Whelan}, {Kerzendorf}, {Conley}, {Crighton},
  {Barbary}, {Muna}, {Ferguson}, {Grollier}, {Parikh}, {Nair}, {Unther},
  {Deil}, {Woillez}, {Conseil}, {Kramer}, {Turner}, {Singer}, {Fox}, {Weaver},
  {Zabalza}, {Edwards}, {Azalee Bostroem}, {Burke}, {Casey}, {Crawford},
  {Dencheva}, {Ely}, {Jenness}, {Labrie}, {Lim}, {Pierfederici}, {Pontzen},
  {Ptak}, {Refsdal}, {Servillat}, \& {Streicher}}]{Astropy2013}
{Astropy Collaboration}, {Robitaille}, T.~P., {Tollerud}, E.~J., {et~al.} 2013,
  \aap, 558, A33, \dodoi{10.1051/0004-6361/201322068}

\bibitem[{{Astropy Collaboration} {et~al.}(2018){Astropy Collaboration},
  {Price-Whelan}, {Sip{\H{o}}cz}, {G{\"u}nther}, {Lim}, {Crawford}, {Conseil},
  {Shupe}, {Craig}, {Dencheva}, {Ginsburg}, {VanderPlas}, {Bradley},
  {P{\'e}rez-Su{\'a}rez}, {de Val-Borro}, {Aldcroft}, {Cruz}, {Robitaille},
  {Tollerud}, {Ardelean}, {Babej}, {Bach}, {Bachetti}, {Bakanov}, {Bamford},
  {Barentsen}, {Barmby}, {Baumbach}, {Berry}, {Biscani}, {Boquien}, {Bostroem},
  {Bouma}, {Brammer}, {Bray}, {Breytenbach}, {Buddelmeijer}, {Burke},
  {Calderone}, {Cano Rodr{\'\i}guez}, {Cara}, {Cardoso}, {Cheedella}, {Copin},
  {Corrales}, {Crichton}, {D'Avella}, {Deil}, {Depagne}, {Dietrich}, {Donath},
  {Droettboom}, {Earl}, {Erben}, {Fabbro}, {Ferreira}, {Finethy}, {Fox},
  {Garrison}, {Gibbons}, {Goldstein}, {Gommers}, {Greco}, {Greenfield},
  {Groener}, {Grollier}, {Hagen}, {Hirst}, {Homeier}, {Horton}, {Hosseinzadeh},
  {Hu}, {Hunkeler}, {Ivezi{\'c}}, {Jain}, {Jenness}, {Kanarek}, {Kendrew},
  {Kern}, {Kerzendorf}, {Khvalko}, {King}, {Kirkby}, {Kulkarni}, {Kumar},
  {Lee}, {Lenz}, {Littlefair}, {Ma}, {Macleod}, {Mastropietro}, {McCully},
  {Montagnac}, {Morris}, {Mueller}, {Mumford}, {Muna}, {Murphy}, {Nelson},
  {Nguyen}, {Ninan}, {N{\"o}the}, {Ogaz}, {Oh}, {Parejko}, {Parley}, {Pascual},
  {Patil}, {Patil}, {Plunkett}, {Prochaska}, {Rastogi}, {Reddy Janga},
  {Sabater}, {Sakurikar}, {Seifert}, {Sherbert}, {Sherwood-Taylor}, {Shih},
  {Sick}, {Silbiger}, {Singanamalla}, {Singer}, {Sladen}, {Sooley},
  {Sornarajah}, {Streicher}, {Teuben}, {Thomas}, {Tremblay}, {Turner},
  {Terr{\'o}n}, {van Kerkwijk}, {de la Vega}, {Watkins}, {Weaver}, {Whitmore},
  {Woillez}, {Zabalza}, \& {Astropy Contributors}}]{Astropy2018}
{Astropy Collaboration}, {Price-Whelan}, A.~M., {Sip{\H{o}}cz}, B.~M., {et~al.}
  2018, \aj, 156, 123, \dodoi{10.3847/1538-3881/aabc4f}

\bibitem[{{Belli} {et~al.}(2014){Belli}, {Newman}, \& {Ellis}}]{Belli14}
{Belli}, S., {Newman}, A.~B., \& {Ellis}, R.~S. 2014, \apj, 783, 117,
  \dodoi{10.1088/0004-637X/783/2/117}

\bibitem[{{Bertin} \& {Arnouts}(1996)}]{Bertin96}
{Bertin}, E., \& {Arnouts}, S. 1996, \aaps, 117, 393,
  \dodoi{10.1051/aas:1996164}

\bibitem[{{Beverage} {et~al.}(2021){Beverage}, {Kriek}, {Conroy}, {Bezanson},
  {Franx}, \& {van der Wel}}]{Beverage21}
{Beverage}, A.~G., {Kriek}, M., {Conroy}, C., {et~al.} 2021, \apjl, 917, L1,
  \dodoi{10.3847/2041-8213/ac12cd}

\bibitem[{{Calura} {et~al.}(2009){Calura}, {Pipino}, {Chiappini}, {Matteucci},
  \& {Maiolino}}]{Calura2009}
{Calura}, F., {Pipino}, A., {Chiappini}, C., {Matteucci}, F., \& {Maiolino}, R.
  2009, \aap, 504, 373, \dodoi{10.1051/0004-6361/200911756}

\bibitem[{{Calzetti} {et~al.}(2000){Calzetti}, {Armus}, {Bohlin}, {Kinney},
  {Koornneef}, \& {Storchi-Bergmann}}]{Calzetti2000}
{Calzetti}, D., {Armus}, L., {Bohlin}, R.~C., {et~al.} 2000, \apj, 533, 682,
  \dodoi{10.1086/308692}

\bibitem[{{Carnall} {et~al.}(2019){Carnall}, {Leja}, {Johnson}, {McLure},
  {Dunlop}, \& {Conroy}}]{Carnall19}
{Carnall}, A.~C., {Leja}, J., {Johnson}, B.~D., {et~al.} 2019, \apj, 873, 44,
  \dodoi{10.3847/1538-4357/ab04a2}

\bibitem[{{Carnall} {et~al.}(2018){Carnall}, {McLure}, {Dunlop}, \&
  {Dav{\'e}}}]{Carnall18}
{Carnall}, A.~C., {McLure}, R.~J., {Dunlop}, J.~S., \& {Dav{\'e}}, R. 2018,
  \mnras, 480, 4379, \dodoi{10.1093/mnras/sty2169}

\bibitem[{{Carnall} {et~al.}(2022){Carnall}, {McLure}, {Dunlop}, {Hamadouche},
  {Cullen}, {McLeod}, {Begley}, {Amorin}, {Bolzonella}, {Castellano},
  {Cimatti}, {Fontanot}, {Gargiulo}, {Garilli}, {Mannucci}, {Pentericci},
  {Talia}, {Zamorani}, {Calabro}, {Cresci}, \& {Hathi}}]{Carnall22}
{Carnall}, A.~C., {McLure}, R.~J., {Dunlop}, J.~S., {et~al.} 2022, \apj, 929,
  131, \dodoi{10.3847/1538-4357/ac5b62}

\bibitem[{{Choi} {et~al.}(2014){Choi}, {Conroy}, {Moustakas}, {Graves},
  {Holden}, {Brodwin}, {Brown}, \& {van Dokkum}}]{Choi14}
{Choi}, J., {Conroy}, C., {Moustakas}, J., {et~al.} 2014, \apj, 792, 95,
  \dodoi{10.1088/0004-637X/792/2/95}

\bibitem[{{Choi} {et~al.}(2016){Choi}, {Dotter}, {Conroy}, {Cantiello},
  {Paxton}, \& {Johnson}}]{Choi2016}
{Choi}, J., {Dotter}, A., {Conroy}, C., {et~al.} 2016, \apj, 823, 102,
  \dodoi{10.3847/0004-637X/823/2/102}

\bibitem[{{Conroy} \& {van Dokkum}(2012)}]{Conroy12}
{Conroy}, C., \& {van Dokkum}, P.~G. 2012, \apj, 760, 71,
  \dodoi{10.1088/0004-637X/760/1/71}

\bibitem[{{Conroy} {et~al.}(2018){Conroy}, {Villaume}, {van Dokkum}, \&
  {Lind}}]{Conroy2018}
{Conroy}, C., {Villaume}, A., {van Dokkum}, P.~G., \& {Lind}, K. 2018, \apj,
  854, 139, \dodoi{10.3847/1538-4357/aaab49}

\bibitem[{{de los Reyes} {et~al.}(2022){de los Reyes}, {Kirby}, {Ji}, \&
  {Nu{\~n}ez}}]{delosReyes22}
{de los Reyes}, M. A.~C., {Kirby}, E.~N., {Ji}, A.~P., \& {Nu{\~n}ez}, E.~H.
  2022, \apj, 925, 66, \dodoi{10.3847/1538-4357/ac332b}

\bibitem[{{Dekel} \& {Silk}(1986)}]{Dekel1986}
{Dekel}, A., \& {Silk}, J. 1986, \apj, 303, 39, \dodoi{10.1086/164050}

\bibitem[{{Dey} {et~al.}(2019){Dey}, {Schlegel}, {Lang}, {Blum}, {Burleigh},
  {Fan}, {Findlay}, {Finkbeiner}, {Herrera}, {Juneau}, {Landriau}, {Levi},
  {McGreer}, {Meisner}, {Myers}, {Moustakas}, {Nugent}, {Patej}, {Schlafly},
  {Walker}, {Valdes}, {Weaver}, {Y{\`e}che}, {Zou}, {Zhou}, {Abareshi},
  {Abbott}, {Abolfathi}, {Aguilera}, {Alam}, {Allen}, {Alvarez}, {Annis},
  {Ansarinejad}, {Aubert}, {Beechert}, {Bell}, {BenZvi}, {Beutler}, {Bielby},
  {Bolton}, {Brice{\~n}o}, {Buckley-Geer}, {Butler}, {Calamida}, {Carlberg},
  {Carter}, {Casas}, {Castander}, {Choi}, {Comparat}, {Cukanovaite}, {Delubac},
  {DeVries}, {Dey}, {Dhungana}, {Dickinson}, {Ding}, {Donaldson}, {Duan},
  {Duckworth}, {Eftekharzadeh}, {Eisenstein}, {Etourneau}, {Fagrelius},
  {Farihi}, {Fitzpatrick}, {Font-Ribera}, {Fulmer}, {G{\"a}nsicke},
  {Gaztanaga}, {George}, {Gerdes}, {Gontcho}, {Gorgoni}, {Green}, {Guy},
  {Harmer}, {Hernandez}, {Honscheid}, {Huang}, {James}, {Jannuzi}, {Jiang},
  {Joyce}, {Karcher}, {Karkar}, {Kehoe}, {Kneib}, {Kueter-Young}, {Lan},
  {Lauer}, {Le Guillou}, {Le Van Suu}, {Lee}, {Lesser}, {Perreault Levasseur},
  {Li}, {Mann}, {Marshall}, {Mart{\'\i}nez-V{\'a}zquez}, {Martini}, {du Mas des
  Bourboux}, {McManus}, {Meier}, {M{\'e}nard}, {Metcalfe},
  {Mu{\~n}oz-Guti{\'e}rrez}, {Najita}, {Napier}, {Narayan}, {Newman}, {Nie},
  {Nord}, {Norman}, {Olsen}, {Paat}, {Palanque-Delabrouille}, {Peng},
  {Poppett}, {Poremba}, {Prakash}, {Rabinowitz}, {Raichoor}, {Rezaie},
  {Robertson}, {Roe}, {Ross}, {Ross}, {Rudnick}, {Safonova}, {Saha},
  {S{\'a}nchez}, {Savary}, {Schweiker}, {Scott}, {Seo}, {Shan}, {Silva},
  {Slepian}, {Soto}, {Sprayberry}, {Staten}, {Stillman}, {Stupak}, {Summers},
  {Sien Tie}, {Tirado}, {Vargas-Maga{\~n}a}, {Vivas}, {Wechsler}, {Williams},
  {Yang}, {Yang}, {Yapici}, {Zaritsky}, {Zenteno}, {Zhang}, {Zhang}, {Zhou}, \&
  {Zhou}}]{Dey19}
{Dey}, A., {Schlegel}, D.~J., {Lang}, D., {et~al.} 2019, \aj, 157, 168,
  \dodoi{10.3847/1538-3881/ab089d}

\bibitem[{{Erb} {et~al.}(2006){Erb}, {Shapley}, {Pettini}, {Steidel}, {Reddy},
  \& {Adelberger}}]{Erb06}
{Erb}, D.~K., {Shapley}, A.~E., {Pettini}, M., {et~al.} 2006, \apj, 644, 813,
  \dodoi{10.1086/503623}

\bibitem[{{Estrada-Carpenter} {et~al.}(2019){Estrada-Carpenter}, {Papovich},
  {Momcheva}, {Brammer}, {Long}, {Quadri}, {Bridge}, {Dickinson}, {Ferguson},
  {Finkelstein}, {Giavalisco}, {Gosmeyer}, {Lotz}, {Salmon}, {Skelton},
  {Trump}, \& {Weiner}}]{Estrada-Carpenter19}
{Estrada-Carpenter}, V., {Papovich}, C., {Momcheva}, I., {et~al.} 2019, \apj,
  870, 133, \dodoi{10.3847/1538-4357/aaf22e}

\bibitem[{{Finlator} \& {Dav{\'e}}(2008)}]{Finlator2008}
{Finlator}, K., \& {Dav{\'e}}, R. 2008, \mnras, 385, 2181,
  \dodoi{10.1111/j.1365-2966.2008.12991.x}

\bibitem[{{Foreman-Mackey} {et~al.}(2013){Foreman-Mackey}, {Hogg}, {Lang}, \&
  {Goodman}}]{Foreman-Mackey2013}
{Foreman-Mackey}, D., {Hogg}, D.~W., {Lang}, D., \& {Goodman}, J. 2013, \pasp,
  125, 306, \dodoi{10.1086/670067}

\bibitem[{{Gallazzi} {et~al.}(2014){Gallazzi}, {Bell}, {Zibetti}, {Brinchmann},
  \& {Kelson}}]{Gallazzi14}
{Gallazzi}, A., {Bell}, E.~F., {Zibetti}, S., {Brinchmann}, J., \& {Kelson},
  D.~D. 2014, \apj, 788, 72, \dodoi{10.1088/0004-637X/788/1/72}

\bibitem[{{Gallazzi} {et~al.}(2005){Gallazzi}, {Charlot}, {Brinchmann},
  {White}, \& {Tremonti}}]{Gallazzi05}
{Gallazzi}, A., {Charlot}, S., {Brinchmann}, J., {White}, S. D.~M., \&
  {Tremonti}, C.~A. 2005, \mnras, 362, 41,
  \dodoi{10.1111/j.1365-2966.2005.09321.x}

\bibitem[{{Greene} {et~al.}(2013){Greene}, {Murphy}, {Graves}, {Gunn},
  {Raskutti}, {Comerford}, \& {Gebhardt}}]{Greene13}
{Greene}, J.~E., {Murphy}, J.~D., {Graves}, G.~J., {et~al.} 2013, \apj, 776,
  64, \dodoi{10.1088/0004-637X/776/2/64}

\bibitem[{Harris {et~al.}(2020)Harris, Millman, van~der Walt, Gommers,
  Virtanen, Cournapeau, Wieser, Taylor, Berg, Smith, Kern, Picus, Hoyer, van
  Kerkwijk, Brett, Haldane, Fernández~del Río, Wiebe, Peterson,
  Gérard-Marchant, Sheppard, Reddy, Weckesser, Abbasi, Gohlke, \&
  Oliphant}]{2020NumPy-Array}
Harris, C.~R., Millman, K.~J., van~der Walt, S.~J., {et~al.} 2020, Nature, 585,
  357–362, \dodoi{10.1038/s41586-020-2649-2}

\bibitem[{{Hayward} \& {Hopkins}(2017)}]{Hayward17}
{Hayward}, C.~C., \& {Hopkins}, P.~F. 2017, \mnras, 465, 1682,
  \dodoi{10.1093/mnras/stw2888}

\bibitem[{{Horne}(1986)}]{Horne86}
{Horne}, K. 1986, \pasp, 98, 609, \dodoi{10.1086/131801}

\bibitem[{{Jacobs} {et~al.}(2019{\natexlab{a}}){Jacobs}, {Collett},
  {Glazebrook}, {McCarthy}, {Qin}, {Abbott}, {Abdalla}, {Annis}, {Avila},
  {Bechtol}, {Bertin}, {Brooks}, {Buckley-Geer}, {Burke}, {Carnero Rosell},
  {Carrasco Kind}, {Carretero}, {da Costa}, {Davis}, {De Vicente}, {Desai},
  {Diehl}, {Doel}, {Eifler}, {Flaugher}, {Frieman}, {Garc{\'\i}a-Bellido},
  {Gaztanaga}, {Gerdes}, {Goldstein}, {Gruen}, {Gruendl}, {Gschwend},
  {Gutierrez}, {Hartley}, {Hollowood}, {Honscheid}, {Hoyle}, {James}, {Kuehn},
  {Kuropatkin}, {Lahav}, {Li}, {Lima}, {Lin}, {Maia}, {Martini}, {Miller},
  {Miquel}, {Nord}, {Plazas}, {Sanchez}, {Scarpine}, {Schubnell}, {Serrano},
  {Sevilla-Noarbe}, {Smith}, {Soares-Santos}, {Sobreira}, {Suchyta}, {Swanson},
  {Tarle}, {Vikram}, {Walker}, {Zhang}, {Zuntz}, \& {DES
  Collaboration}}]{Jacobs19a}
{Jacobs}, C., {Collett}, T., {Glazebrook}, K., {et~al.} 2019{\natexlab{a}},
  \mnras, 484, 5330, \dodoi{10.1093/mnras/stz272}

\bibitem[{{Jacobs} {et~al.}(2019{\natexlab{b}}){Jacobs}, {Collett},
  {Glazebrook}, {Buckley-Geer}, {Diehl}, {Lin}, {McCarthy}, {Qin}, {Odden},
  {Caso Escudero}, {Dial}, {Yung}, {Gaitsch}, {Pellico}, {Lindgren}, {Abbott},
  {Annis}, {Avila}, {Brooks}, {Burke}, {Carnero Rosell}, {Carrasco Kind},
  {Carretero}, {da Costa}, {De Vicente}, {Fosalba}, {Frieman},
  {Garc{\'\i}a-Bellido}, {Gaztanaga}, {Goldstein}, {Gruen}, {Gruendl},
  {Gschwend}, {Hollowood}, {Honscheid}, {Hoyle}, {James}, {Krause},
  {Kuropatkin}, {Lahav}, {Lima}, {Maia}, {Marshall}, {Miquel}, {Plazas},
  {Roodman}, {Sanchez}, {Scarpine}, {Serrano}, {Sevilla-Noarbe}, {Smith},
  {Sobreira}, {Suchyta}, {Swanson}, {Tarle}, {Vikram}, {Walker}, {Zhang}, \&
  {DES Collaboration}}]{Jacobs19b}
---. 2019{\natexlab{b}}, \apjs, 243, 17, \dodoi{10.3847/1538-4365/ab26b6}

\bibitem[{{Jafariyazani} {et~al.}(2020){Jafariyazani}, {Newman}, {Mobasher},
  {Belli}, {Ellis}, \& {Patel}}]{Jafariyazani20}
{Jafariyazani}, M., {Newman}, A.~B., {Mobasher}, B., {et~al.} 2020, \apjl, 897,
  L42, \dodoi{10.3847/2041-8213/aba11c}

\bibitem[{{Johnson} {et~al.}(2021){Johnson}, {Leja}, {Conroy}, \&
  {Speagle}}]{Johnson21}
{Johnson}, B.~D., {Leja}, J., {Conroy}, C., \& {Speagle}, J.~S. 2021, \apjs,
  254, 22, \dodoi{10.3847/1538-4365/abef67}

\bibitem[{{Jullo} {et~al.}(2007){Jullo}, {Kneib}, {Limousin},
  {El{\'\i}asd{\'o}ttir}, {Marshall}, \& {Verdugo}}]{Jullo07}
{Jullo}, E., {Kneib}, J.~P., {Limousin}, M., {et~al.} 2007, New Journal of
  Physics, 9, 447, \dodoi{10.1088/1367-2630/9/12/447}

\bibitem[{{Kacprzak} {et~al.}(2016){Kacprzak}, {van de Voort}, {Glazebrook},
  {Tran}, {Yuan}, {Nanayakkara}, {Allen}, {Alcorn}, {Cowley}, {Labb{\'e}},
  {Spitler}, {Straatman}, \& {Tomczak}}]{Kacprzak16}
{Kacprzak}, G.~G., {van de Voort}, F., {Glazebrook}, K., {et~al.} 2016, \apjl,
  826, L11, \dodoi{10.3847/2041-8205/826/1/L11}

\bibitem[{{Kirby} {et~al.}(2013){Kirby}, {Cohen}, {Guhathakurta}, {Cheng},
  {Bullock}, \& {Gallazzi}}]{Kirby2013}
{Kirby}, E.~N., {Cohen}, J.~G., {Guhathakurta}, P., {et~al.} 2013, \apj, 779,
  102, \dodoi{10.1088/0004-637X/779/2/102}

\bibitem[{{Kriek} {et~al.}(2016){Kriek}, {Conroy}, {van Dokkum}, {Shapley},
  {Choi}, {Reddy}, {Siana}, {van de Voort}, {Coil}, \& {Mobasher}}]{Kriek16}
{Kriek}, M., {Conroy}, C., {van Dokkum}, P.~G., {et~al.} 2016, \nat, 540, 248,
  \dodoi{10.1038/nature20570}

\bibitem[{{Kriek} {et~al.}(2019){Kriek}, {Price}, {Conroy}, {Suess}, {Mowla},
  {Pasha}, {Bezanson}, {van Dokkum}, \& {Barro}}]{Kriek19}
{Kriek}, M., {Price}, S.~H., {Conroy}, C., {et~al.} 2019, \apjl, 880, L31,
  \dodoi{10.3847/2041-8213/ab2e75}

\bibitem[{{Kroupa}(2001)}]{Kroupa2001}
{Kroupa}, P. 2001, \mnras, 322, 231, \dodoi{10.1046/j.1365-8711.2001.04022.x}

\bibitem[{{Leethochawalit} {et~al.}(2016){Leethochawalit}, {Jones}, {Ellis},
  {Stark}, {Richard}, {Zitrin}, \& {Auger}}]{Leethochawalit16}
{Leethochawalit}, N., {Jones}, T.~A., {Ellis}, R.~S., {et~al.} 2016, \apj, 820,
  84, \dodoi{10.3847/0004-637X/820/2/84}

\bibitem[{{Leethochawalit} {et~al.}(2019){Leethochawalit}, {Kirby}, {Ellis},
  {Moran}, \& {Treu}}]{Leethochawalit2019}
{Leethochawalit}, N., {Kirby}, E.~N., {Ellis}, R.~S., {Moran}, S.~M., \&
  {Treu}, T. 2019, \apj, 885, 100, \dodoi{10.3847/1538-4357/ab4809}

\bibitem[{{Leethochawalit} {et~al.}(2018){Leethochawalit}, {Kirby}, {Moran},
  {Ellis}, \& {Treu}}]{Leethochawalit2018}
{Leethochawalit}, N., {Kirby}, E.~N., {Moran}, S.~M., {Ellis}, R.~S., \&
  {Treu}, T. 2018, \apj, 856, 15, \dodoi{10.3847/1538-4357/aab26a}

\bibitem[{{Lonoce} {et~al.}(2015){Lonoce}, {Longhetti}, {Maraston}, {Thomas},
  {Mancini}, {Cimatti}, {Ciocca}, {Citro}, {Daddi}, {di Serego Alighieri},
  {Gargiulo}, {Maiolino}, {Mannucci}, {Moresco}, {Pozzetti}, {Quai}, \&
  {Saracco}}]{Lonoce15}
{Lonoce}, I., {Longhetti}, M., {Maraston}, C., {et~al.} 2015, \mnras, 454,
  3912, \dodoi{10.1093/mnras/stv2150}

\bibitem[{{Lu} {et~al.}(2015){Lu}, {Blanc}, \& {Benson}}]{Lu15}
{Lu}, Y., {Blanc}, G.~A., \& {Benson}, A. 2015, \apj, 808, 129,
  \dodoi{10.1088/0004-637X/808/2/129}

\bibitem[{{Ludlow} {et~al.}(2016){Ludlow}, {Bose}, {Angulo}, {Wang},
  {Hellwing}, {Navarro}, {Cole}, \& {Frenk}}]{Ludlow2016}
{Ludlow}, A.~D., {Bose}, S., {Angulo}, R.~E., {et~al.} 2016, \mnras, 460, 1214,
  \dodoi{10.1093/mnras/stw1046}

\bibitem[{{Lupton} {et~al.}(2004){Lupton}, {Blanton}, {Fekete}, {Hogg},
  {O'Mullane}, {Szalay}, \& {Wherry}}]{Lupton2004}
{Lupton}, R., {Blanton}, M.~R., {Fekete}, G., {et~al.} 2004, \pasp, 116, 133,
  \dodoi{10.1086/382245}

\bibitem[{{Ma} {et~al.}(2016){Ma}, {Hopkins}, {Faucher-Gigu{\`e}re}, {Zolman},
  {Muratov}, {Kere{\v{s}}}, \& {Quataert}}]{Ma16}
{Ma}, X., {Hopkins}, P.~F., {Faucher-Gigu{\`e}re}, C.-A., {et~al.} 2016,
  \mnras, 456, 2140, \dodoi{10.1093/mnras/stv2659}

\bibitem[{{Magrini} {et~al.}(2012){Magrini}, {Hunt}, {Galli}, {Schneider},
  {Bianchi}, {Maiolino}, {Romano}, {Tosi}, \& {Valiante}}]{Magrini2012}
{Magrini}, L., {Hunt}, L., {Galli}, D., {et~al.} 2012, \mnras, 427, 1075,
  \dodoi{10.1111/j.1365-2966.2012.22055.x}

\bibitem[{{Man} {et~al.}(2021){Man}, {Zabl}, {Brammer}, {Richard}, {Toft},
  {Stockmann}, {Gallazzi}, {Zibetti}, \& {Ebeling}}]{Man21}
{Man}, A. W.~S., {Zabl}, J., {Brammer}, G.~B., {et~al.} 2021, \apj, 919, 20,
  \dodoi{10.3847/1538-4357/ac0ae3}

\bibitem[{{Mendel} {et~al.}(2020){Mendel}, {Beifiori}, {Saglia}, {Bender},
  {Brammer}, {Chan}, {F{\"o}rster Schreiber}, {Fossati}, {Galametz},
  {Momcheva}, {Nelson}, {Wilman}, \& {Wuyts}}]{Mendel2020}
{Mendel}, J.~T., {Beifiori}, A., {Saglia}, R.~P., {et~al.} 2020, \apj, 899, 87,
  \dodoi{10.3847/1538-4357/ab9ffc}

\bibitem[{{Morishita} {et~al.}(2018){Morishita}, {Abramson}, {Treu}, {Wang},
  {Brammer}, {Kelly}, {Stiavelli}, {Jones}, {Schmidt}, {Trenti}, \&
  {Vulcani}}]{Morishita18}
{Morishita}, T., {Abramson}, L.~E., {Treu}, T., {et~al.} 2018, \apjl, 856, L4,
  \dodoi{10.3847/2041-8213/aab493}

\bibitem[{{Nanayakkara} {et~al.}(2022){Nanayakkara}, {Esdaile}, {Glazebrook},
  {Espejo Salcedo}, {Durre}, \& {Jacobs}}]{Nanayakkara22}
{Nanayakkara}, T., {Esdaile}, J., {Glazebrook}, K., {et~al.} 2022, \pasa, 39,
  e002, \dodoi{10.1017/pasa.2021.61}

\bibitem[{{Navarro} {et~al.}(1997){Navarro}, {Frenk}, \& {White}}]{nfw}
{Navarro}, J.~F., {Frenk}, C.~S., \& {White}, S. D.~M. 1997, \apj, 490, 493,
  \dodoi{10.1086/304888}

\bibitem[{{Newman} {et~al.}(2018){Newman}, {Belli}, {Ellis}, \&
  {Patel}}]{Newman18}
{Newman}, A.~B., {Belli}, S., {Ellis}, R.~S., \& {Patel}, S.~G. 2018, \apj,
  862, 126, \dodoi{10.3847/1538-4357/aacd4f}

\bibitem[{Nightingale \& Dye(2015)}]{Nightingale2015}
Nightingale, J.~W., \& Dye, S. 2015, MNRAS, 452, 2940,
  \dodoi{10.1093/mnras/stv1455}

\bibitem[{Nightingale {et~al.}(2018)Nightingale, Dye, \&
  Massey}]{Nightingale2018}
Nightingale, J.~W., Dye, S., \& Massey, R.~J. 2018, MNRAS, 478, 4738,
  \dodoi{10.1093/mnras/sty1264}

\bibitem[{Nightingale {et~al.}(2021{\natexlab{a}})Nightingale, Hayes, \&
  Griffiths}]{pyautofit}
Nightingale, J.~W., Hayes, R.~G., \& Griffiths, M. 2021{\natexlab{a}}, Journal
  of Open Source Software, 6, 2550, \dodoi{10.21105/joss.02550}

\bibitem[{Nightingale {et~al.}(2021{\natexlab{b}})Nightingale, Hayes, Kelly,
  Amvrosiadis, Etherington, He, Li, Cao, Frawley, Cole, Enia, Frenk, Harvey,
  Li, Massey, Negrello, \& Robertson}]{pyautolens}
Nightingale, J.~W., Hayes, R.~G., Kelly, A., {et~al.} 2021{\natexlab{b}},
  Journal of Open Source Software, 6, 2825, \dodoi{10.21105/joss.02825}

\bibitem[{{Nomoto} {et~al.}(2006){Nomoto}, {Tominaga}, {Umeda}, {Kobayashi}, \&
  {Maeda}}]{Nomoto06}
{Nomoto}, K., {Tominaga}, N., {Umeda}, H., {Kobayashi}, C., \& {Maeda}, K.
  2006, \nphysa, 777, 424, \dodoi{10.1016/j.nuclphysa.2006.05.008}

\bibitem[{{Onodera} {et~al.}(2015){Onodera}, {Carollo}, {Renzini},
  {Cappellari}, {Mancini}, {Arimoto}, {Daddi}, {Gobat}, {Strazzullo},
  {Tacchella}, \& {Yamada}}]{Onodera15}
{Onodera}, M., {Carollo}, C.~M., {Renzini}, A., {et~al.} 2015, \apj, 808, 161,
  \dodoi{10.1088/0004-637X/808/2/161}

\bibitem[{{Oser} {et~al.}(2010){Oser}, {Ostriker}, {Naab}, {Johansson}, \&
  {Burkert}}]{Oser2010}
{Oser}, L., {Ostriker}, J.~P., {Naab}, T., {Johansson}, P.~H., \& {Burkert}, A.
  2010, \apj, 725, 2312, \dodoi{10.1088/0004-637X/725/2/2312}

\bibitem[{{Oyarz{\'u}n} {et~al.}(2019){Oyarz{\'u}n}, {Bundy}, {Westfall},
  {Belfiore}, {Thomas}, {Maraston}, {Lian}, {Arag{\'o}n-Salamanca}, {Zheng},
  {Gonzalez-Perez}, {Law}, {Drory}, \& {Andrews}}]{Oyarzun19}
{Oyarz{\'u}n}, G.~A., {Bundy}, K., {Westfall}, K.~B., {et~al.} 2019, \apj, 880,
  111, \dodoi{10.3847/1538-4357/ab297c}

\bibitem[{{Prochaska} {et~al.}(2020{\natexlab{a}}){Prochaska}, {Hennawi},
  {Westfall}, {Cooke}, {Wang}, {Hsyu}, {Davies}, {Farina}, \&
  {Pelliccia}}]{pypeit:joss}
{Prochaska}, J., {Hennawi}, J., {Westfall}, K., {et~al.} 2020{\natexlab{a}},
  The Journal of Open Source Software, 5, 2308, \dodoi{10.21105/joss.02308}

\bibitem[{{Prochaska} {et~al.}(2020{\natexlab{b}}){Prochaska}, {Hennawi},
  {Cooke}, {Westfall}, {Wang}, {EmAstro}, {Tiffanyhsyu}, {Wasserman},
  {Villaume}, {Marijana777}, {Schindler}, {Young}, {Simha}, {Wilde}, {Tejos},
  {Isbell}, {Fl{\"o}rs}, {Sandford}, {Vasovi{\'c}}, {Betts}, \&
  {Holden}}]{pypeit:zenodo}
{Prochaska}, J.~X., {Hennawi}, J., {Cooke}, R., {et~al.} 2020{\natexlab{b}},
  {pypeit/PypeIt: Release 1.0.0}, v1.0.0,  Zenodo,
  \dodoi{10.5281/zenodo.3743493}

\bibitem[{{S{\'a}nchez-Bl{\'a}zquez} {et~al.}(2006){S{\'a}nchez-Bl{\'a}zquez},
  {Peletier}, {Jim{\'e}nez-Vicente}, {Cardiel}, {Cenarro},
  {Falc{\'o}n-Barroso}, {Gorgas}, {Selam}, \& {Vazdekis}}]{Sanchez-Blazquez06}
{S{\'a}nchez-Bl{\'a}zquez}, P., {Peletier}, R.~F., {Jim{\'e}nez-Vicente}, J.,
  {et~al.} 2006, \mnras, 371, 703, \dodoi{10.1111/j.1365-2966.2006.10699.x}

\bibitem[{{Schlafly} \& {Finkbeiner}(2011)}]{Schlafly2011}
{Schlafly}, E.~F., \& {Finkbeiner}, D.~P. 2011, \apj, 737, 103,
  \dodoi{10.1088/0004-637X/737/2/103}

\bibitem[{{Shajib} {et~al.}(2022{\natexlab{a}}){Shajib}, {Glazebrook},
  {Barone}, {Lewis}, {Jones}, {Tran}, {Buckley-Geer}, {Collett}, {Frieman}, \&
  {Jacobs}}]{Shajib22}
{Shajib}, A.~J., {Glazebrook}, K., {Barone}, T., {et~al.} 2022{\natexlab{a}},
  \apj, 938, 141, \dodoi{10.3847/1538-4357/ac927b}

\bibitem[{{Shajib} {et~al.}(2022{\natexlab{b}}){Shajib}, {Glazebrook},
  {Barone}, {Lewis}, {Jones}, {Tran}, {Buckley-Geer}, {Collett}, {Frieman}, \&
  {Jacobs}}]{Shajib22_soft}
---. 2022{\natexlab{b}}, {LensingETC: Lensing Exposure Time Calculator},
  Astrophysics Source Code Library, record ascl:2210.027.
\newblock \doeprint{2210.027}

\bibitem[{{Skrutskie} {et~al.}(2006){Skrutskie}, {Cutri}, {Stiening},
  {Weinberg}, {Schneider}, {Carpenter}, {Beichman}, {Capps}, {Chester},
  {Elias}, {Huchra}, {Liebert}, {Lonsdale}, {Monet}, {Price}, {Seitzer},
  {Jarrett}, {Kirkpatrick}, {Gizis}, {Howard}, {Evans}, {Fowler}, {Fullmer},
  {Hurt}, {Light}, {Kopan}, {Marsh}, {McCallon}, {Tam}, {Van Dyk}, \&
  {Wheelock}}]{Skrutskie06}
{Skrutskie}, M.~F., {Cutri}, R.~M., {Stiening}, R., {et~al.} 2006, \aj, 131,
  1163, \dodoi{10.1086/498708}

\bibitem[{Speagle(2020)}]{dynesty}
Speagle, J.~S. 2020, MNRAS, 493, 3132, \dodoi{10.1093/mnras/staa278}

\bibitem[{{Strom} {et~al.}(2017){Strom}, {Steidel}, {Rudie}, {Trainor},
  {Pettini}, \& {Reddy}}]{Strom2017}
{Strom}, A.~L., {Steidel}, C.~C., {Rudie}, G.~C., {et~al.} 2017, \apj, 836,
  164, \dodoi{10.3847/1538-4357/836/2/164}

\bibitem[{{Sukay} {et~al.}(2022){Sukay}, {Khullar}, {Gladders}, {Sharon},
  {Mahler}, {Napier}, {Bleem}, {Dahle}, {Florian}, {Gozman}, {Lin}, {Martinez},
  {Matthews Acu{\~n}a}, {Medina}, {Merz}, {Sanchez}, {Sisco}, {Kavin Stein},
  {Tavangar}, \& {Whitaker}}]{Sukay22}
{Sukay}, E., {Khullar}, G., {Gladders}, M.~D., {et~al.} 2022, arXiv e-prints,
  arXiv:2203.11957.
\newblock \doarXiv{2203.11957}

\bibitem[{{Thomas} {et~al.}(2003){Thomas}, {Maraston}, \& {Bender}}]{Thomas03}
{Thomas}, D., {Maraston}, C., \& {Bender}, R. 2003, \mnras, 343, 279,
  \dodoi{10.1046/j.1365-8711.2003.06659.x}

\bibitem[{{Thomas} {et~al.}(2005){Thomas}, {Maraston}, {Bender}, \& {Mendes de
  Oliveira}}]{Thomas2005}
{Thomas}, D., {Maraston}, C., {Bender}, R., \& {Mendes de Oliveira}, C. 2005,
  \apj, 621, 673, \dodoi{10.1086/426932}

\bibitem[{{Toft} {et~al.}(2012){Toft}, {Gallazzi}, {Zirm}, {Wold}, {Zibetti},
  {Grillo}, \& {Man}}]{Toft12}
{Toft}, S., {Gallazzi}, A., {Zirm}, A., {et~al.} 2012, \apj, 754, 3,
  \dodoi{10.1088/0004-637X/754/1/3}

\bibitem[{{Tran} {et~al.}(2022){Tran}, {Harshan}, {Glazebrook}, {Keerthi
  Vasan}, {Jones}, {Jacobs}, {Kacprzak}, {Barone}, {Collett}, {Gupta},
  {Henderson}, {Kewley}, {Lopez}, {Nanayakkara}, {Sanders}, \&
  {Sweet}}]{Tran22}
{Tran}, K.-V.~H., {Harshan}, A., {Glazebrook}, K., {et~al.} 2022, \aj, 164,
  148, \dodoi{10.3847/1538-3881/ac7da2}

\bibitem[{{Tremonti} {et~al.}(2004){Tremonti}, {Heckman}, {Kauffmann},
  {Brinchmann}, {Charlot}, {White}, {Seibert}, {Peng}, {Schlegel}, {Uomoto},
  {Fukugita}, \& {Brinkmann}}]{Tremonti2004}
{Tremonti}, C.~A., {Heckman}, T.~M., {Kauffmann}, G., {et~al.} 2004, \apj, 613,
  898, \dodoi{10.1086/423264}

\bibitem[{{Villaume} {et~al.}(2017){Villaume}, {Conroy}, {Johnson}, {Rayner},
  {Mann}, \& {van Dokkum}}]{Villaume17}
{Villaume}, A., {Conroy}, C., {Johnson}, B., {et~al.} 2017, \apjs, 230, 23,
  \dodoi{10.3847/1538-4365/aa72ed}

\bibitem[{Virtanen {et~al.}(2020)Virtanen, Gommers, Oliphant, Haberland, Reddy,
  Cournapeau, Burovski, Peterson, Weckesser, Bright, {van der Walt}, Brett,
  Wilson, Millman, Mayorov, Nelson, Jones, Kern, Larson, Carey, Polat, Feng,
  Moore, {VanderPlas}, Laxalde, Perktold, Cimrman, Henriksen, Quintero, Harris,
  Archibald, Ribeiro, Pedregosa, {van Mulbregt}, \& {SciPy 1.0
  Contributors}}]{2020SciPy-NMeth}
Virtanen, P., Gommers, R., Oliphant, T.~E., {et~al.} 2020, Nature Methods, 17,
  261, \dodoi{10.1038/s41592-019-0686-2}

\bibitem[{{Walcher} {et~al.}(2015){Walcher}, {Coelho}, {Gallazzi}, {Bruzual},
  {Charlot}, \& {Chiappini}}]{Walcher2015}
{Walcher}, C.~J., {Coelho}, P.~R.~T., {Gallazzi}, A., {et~al.} 2015, \aap, 582,
  A46, \dodoi{10.1051/0004-6361/201525924}

\bibitem[{{Zhuang} {et~al.}(2021){Zhuang}, {Kirby}, {Leethochawalit}, \& {de
  los Reyes}}]{Zhuang2021}
{Zhuang}, Z., {Kirby}, E.~N., {Leethochawalit}, N., \& {de los Reyes}, M. A.~C.
  2021, \apj, 920, 63, \dodoi{10.3847/1538-4357/ac1340}

\end{thebibliography}
\bibliographystyle{aasjournal}



\end{document}